\documentclass[aps,prb,reprint,superscriptaddress,showpacs]{revtex4-1}
\usepackage{amsmath,amssymb,bm}
\usepackage{graphicx}
\usepackage{epstopdf}
\usepackage{latexsym}
\usepackage{subfigure}
\usepackage{color}
\usepackage{hyperref}

\newcommand{\hamilt}{{\widehat{\mathcal{H}}}}
\newcommand{\spop}{\widehat{\mathbf{S}}}
\newcommand{\spin}{\widehat{S}}
\newcommand{\aver}[1]{\langle #1 \rangle}
\newcommand{\Cupz}{Cu(pz)$_{2}$(ClO$_{4}$)$_{2}$}

\begin{document}

\title{ Switching  of anisotropy and phase diagram\\
 of a Heisenberg square lattice $S=1/2$ antiferromagnet \Cupz.}

\author{K. Yu. Povarov}
    \affiliation{P. L. Kapitza Institute for Physical Problems, RAS, 119334 Moscow, Russia}

\author{A. I. Smirnov}
    \affiliation{P. L. Kapitza Institute for Physical Problems, RAS, 119334 Moscow, Russia}
    \affiliation{Moscow Institute for Physics and Technology , 141700, Dolgoprudny, Russia}

\author{C. P. Landee}
    \affiliation{Department of Physics, Clark University, Worcester, Massachusetts 01610, USA}

\date{\today}

\begin{abstract}
Experiments in the  antiferromagnetic phase of a quasi 2D  $S=1/2$ quasi-square lattice antiferromagnet \Cupz\
reveal a biaxial type of the anisotropy, instead of  the easy-plane one, considered before. The weak in-plane
anisotropy, found by means of electron spin resonance spectroscopy and magnetization measurements, is about an
order of magnitude weaker, than the off-plane anisotropy. The weak in-plane anisotropy results in a spin-flop
phase transition for the magnetic field aligned along easy axis, and, thereby, in a bicritical point on the
phase diagram.  A remarkable feature of the weak in-plane anisotropy is the abrupt change of its sign at the
spin-flop point. This anisotropy switching disappears at the tilting of magnetic field to the easy axis by the
angle of 10$^\circ$ within the plane. The nature of the abrupt anisotropy reversal remains unclear. The phase
diagram is characterized by the increase of the ordering temperature $T_{N}$ in the magnetic field used, except
for a dip near the bicritical point.
\end{abstract}


\maketitle

\section{Introduction \label{Introduction}}

\begin{figure}
  \includegraphics[width=0.45\textwidth]{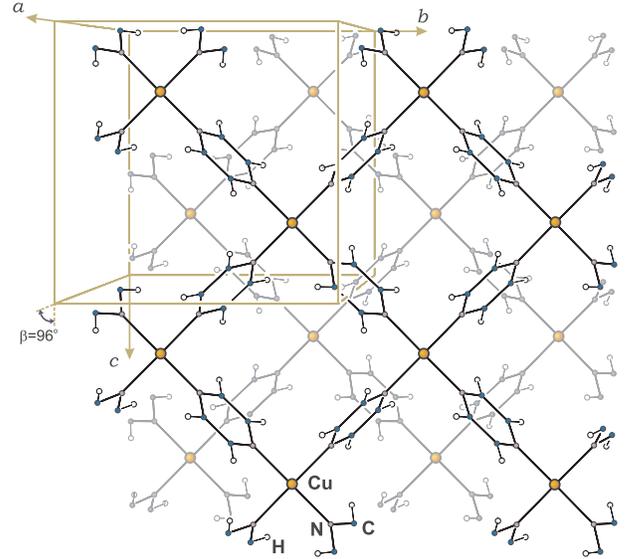}\\
  \caption{(Color online) \Cupz\ structure. Two layers, each containing a square magnetic lattice,
  are displayed; ClO$_{4}$ complexes are not shown for
  clarity. Colors of ions and connecting lines in the lower layer are faded out.
    Crystallographic data are taken from [\onlinecite{chemistry}].}\label{FIG:directions}
\end{figure}

Heisenberg $S=1/2$ antiferromagnet on a square lattice (HSLAF) is a
popular model of low-dimensional magnetism\cite{Interest}. An ideal
HSLAF has no long-range order except at $T=0$, where a N\'{e}el-type
ground state with 40\% reduction of ordered spin component should be
realized\cite{SQLnum}. In real \emph{quasi}-2D antiferromagnet a
weak interlayer interaction is present, providing a N\'{e}el order
at $T>0$. An organometallic compound \Cupz\ (copper pyrazine
perchlorate) has been considered as an example of quasi-2D HSLAF:
copper ions carrying spin $S=1/2$ are bridged together in a slightly
distorted square lattice layers by pyrazine (C$_4$H$_4$N$_2$) rings
as shown on Fig. \ref{FIG:directions}.

Copper pyrazine perchlorate crystallizes from a solution in the
space group $C2/m$ at room temperature, however, at cooling,  near
180 K, there is a phase transition to a structure which has the
space group $C2/c$  [\onlinecite{chemistry}]. In the low temperature
phase, parameters $b$ and $c$ of the monoclinic lattice are close to
each other. The rectangles with sides $b$ and $c$ are approximately
squares. Diagonals of these rectangles form a rhombic lattice, these
rhombuses are slightly distorted squares. The angle between the $a$
and $c$ axes differs only a little from $90^\circ$, thus the lattice
may be represented as weakly distorted tetragonal lattice. Magnetic
ions Cu$^{2+}$ ($S$=1/2), placed at the corners of rhombuses, form
layers in $bc$ planes. The nearest neighbor exchange paths  are
symmetrically equivalent and exactly identical. Therefore the
exchange network within the $bc$ planes is equivalent to that of a
square lattice. Due to ClO$_4$ complexes between the layers, they
are well separated, as well as due to half a period in-plane shift
between layers. Because of this shift a magnetic ion within a layer
is equidistant from four ions in adjacent layer, therefore the
interlayer coupling is canceled in the first order\cite{chemistry}.

Indeed, estimation of the interlayer effective exchange $J_{\perp}$
from values of $T_{N}=4.25$ K and nearest-neighbor exchange $J=18.1$
K by an empirical relation
$J_{\perp}/J\sim\exp{(-\frac{2.3J}{T_{N}})}$ derived from quantum
Monte-Carlo simulation\cite{interplane}, leads to a very small value
of $J_{\perp}\simeq9\cdot10^{-4}J$ [\onlinecite{muons}]. Magnetic
moment per Cu$^{2+}$ ion in the two sublattice structure is only
$0.47\mu_{B}$ at $T\rightarrow0$ in zero field, as detected by
elastic neutron scattering\cite{domain}. This quantum spin reduction
indicates strong influence of quantum fluctuations on the ground
state. From the observation of increase of ordered spin component in
external field Tsyrulin \textit{et al}\cite{domain} conclude, that
the fluctuations are suppressed by magnetic field.  A related
evidence of fluctuations suppression is the significant growth of
$T_{N}$ in magnetic field, confirmed by neutron scattering and
specific heat measurements\cite{domain}. A gap of $E_{0}\simeq0.2$
meV in the spin-wave spectrum, detected by inelastic neutron
scattering\cite{domain,neutrons}, was ascribed to easy-plane
($XY$-type) anisotropy, keeping the spins within the $bc$ plane. The
observation of minimum in susceptibility vs temperature dependence
for a field directed perpendicular to $bc$ plane is consistent with
quantum Monte-Carlo (QMC) simulations\cite{XYcucolliPRL}, which
predict the minimum of the susceptibility for HSLAF with a small
$XY$ anisotropy.

We describe systematic investigations of \Cupz\ by means of multifrequency electron spin resonance (ESR)
spectroscopy and magnetization measurements for different orientations of magnetic field. Our main result is the
observation and measurement of a weak in-plane anisotropy, not detected in previous measurements. This weak
anisotropy induces remarkable features of the phase diagram. These are i) the spin-flop phase transition in a
magnetic field applied along the easy axis, ii) a bicritical point and iii) a dip in $T_{N}(H)$ dependence near
the bicritical point. From antiferromagnetic resonance spectrum we also find that the weak in-plane anisotropy
is surprisingly changing its sign by a jump at the spin-flop point. Besides, this effect of abrupt anisotropy
reversal arises as another phase transition at tilting the magnetic field, at a critical angle between the
magnetic field and the easy axis.

\section{Experiment \label{SECTIONExperiment}}

Samples of \Cupz\ have been grown in Clark University as described
in [\onlinecite{chemistry}]. The lattice parameters of the
monoclinic $C2/c$ lattice are $a=14.072(5)$, $b=9.786(3)$ and
$c=9.781(3)$ {\AA}; $\beta=96.458(4)^{\circ}$. Crystals are flat
rectangular plaquettes colored blue, of the typical  size of 2x2
mm$^2$; the plane of square lattice ($bc$ plane) coincides with the
plane of the plaquette. Sides of  square crystal plaquettes are
aligned at $45^{\circ}$ to $b$ and $c$ axes, coinciding with the
directing lines of magnetic square lattice, see a sketch on margins
of Fig.\ref{FIG:rosettes}.

ESR experiments were performed in Kapitza Institute, using a set of resonator spectrometric inserts in $^4$He
pumping cryostat with a cryomagnet. The frequency range from $5$ to $140$ GHz was covered. A spectrometric
insert for $18-140$ GHz range has a rotable sample holder, allowing to change the orientation of the sample with
respect to magnetic field during the experiment. A small amount of DPPH, free radical compound with $g=2.00$,
was used as a magnetic field label\cite{dpph}.

Magnetization experiments were performed at the Department of Low
Temperature Physics and Superconductivity of M. V. Lomonosov Moscow
State University on Quantum Design 9 Tesla PPMS machine equipped
with vibrating sample magnetometer (VSM) and at the Neutron
Scattering and Magnetism Group in the Laboratory for Solid State
Physics at ETH Z\"{u}rich with the identical machine. The lowest
available temperature was 1.8 K.

\section{ESR data \label{SECTIONResultsESR}}

\subsection{Temperature evolution of ESR signal \label{SECTIONResultsESR_Tdep}}

\begin{figure}
    \includegraphics[width=0.5\textwidth]{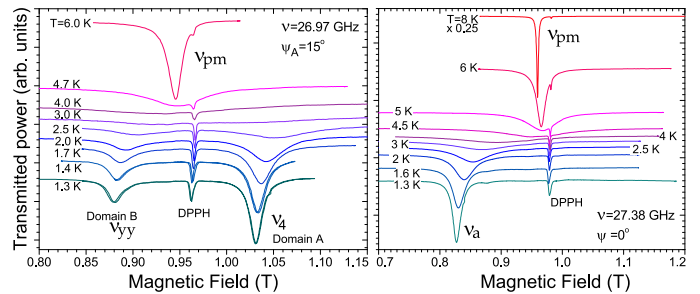}\\
  \caption{(Color online) Left panel: resonance line temperature evolution in a sample, containing two types of domains, when $\psi_{A}=15^{\circ}$.
  Right panel: resonance line temperature evolution in single domain sample when $\psi=0^{\circ}$. In both cases $\xi=0^{\circ}$, i. e. field lies in $xy$ plane.
  The angles $\psi$ and $\xi$ are defined on Fig.\ref{FIG:rosettes}.
  A scaling factor of 0.25
  is applied to 8 K line on the right panel.}\label{FIG:heating}
\end{figure}

\begin{figure}
    \includegraphics[width=0.45\textwidth]{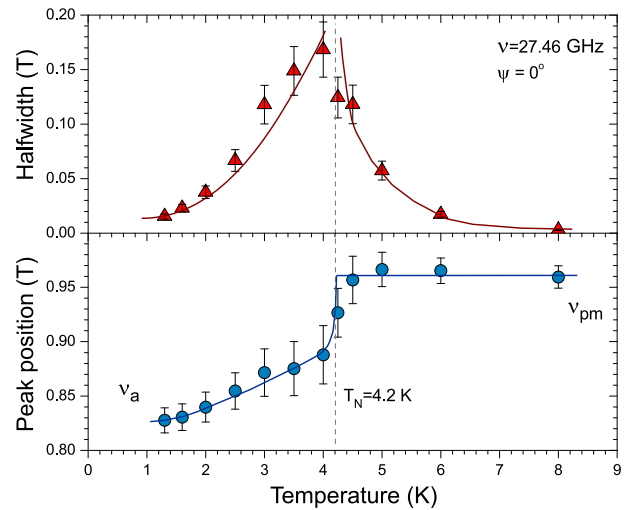}\\
  \caption{(Color online) Changes of resonance field and linewidth through $T_{N}$ in a single domain sample for $\mathbf{H} \parallel x$
   at $\nu=27.46$ GHz; corresponding set of ESR signals is shown at the right panel of Fig. \ref{FIG:heating}.
   Lines are guide to the eye.}\label{FIG:TevolPoints}
\end{figure}

\begin{figure}
  \includegraphics[width=0.5\textwidth]{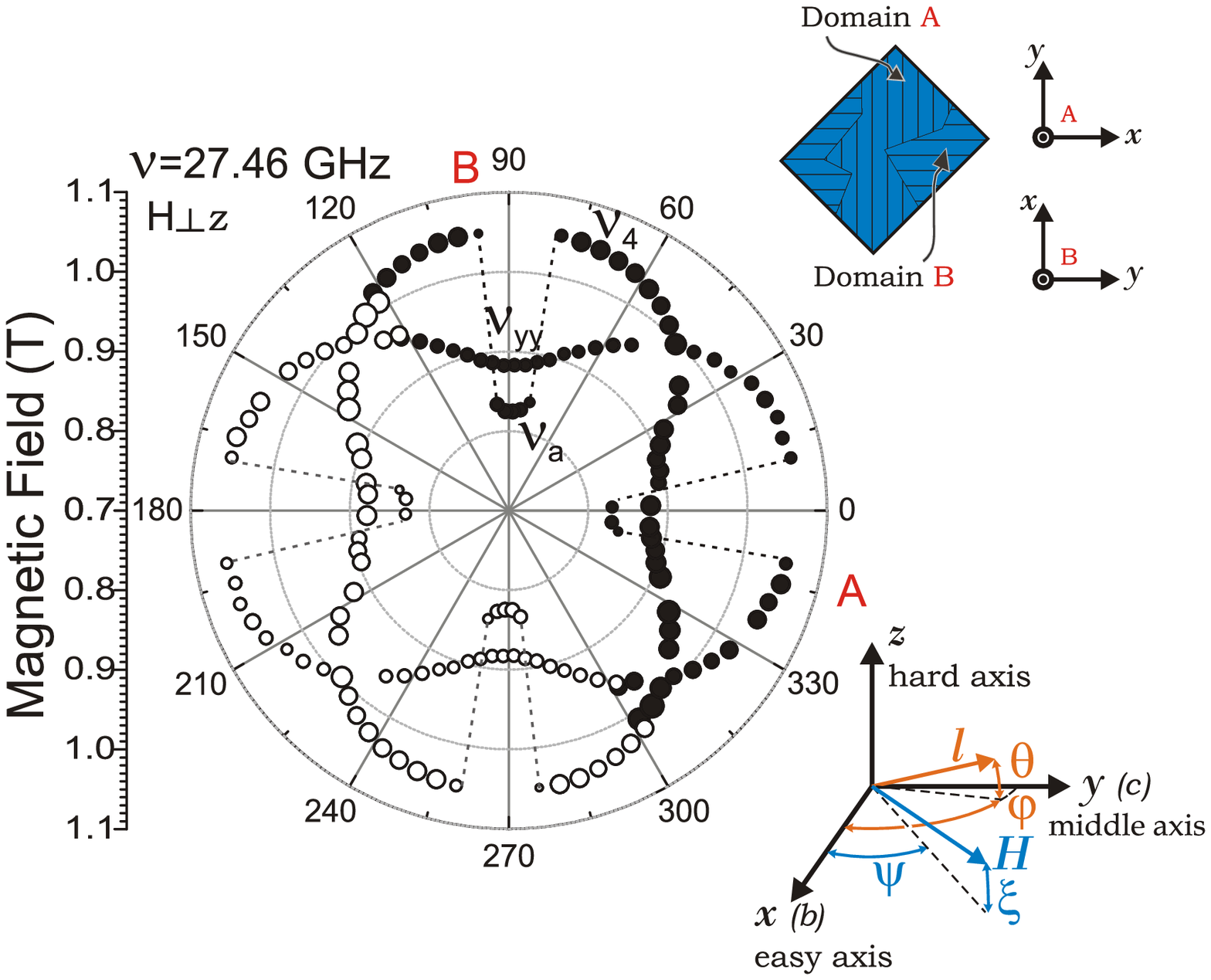}\\
  \caption{(Color online) Polar plot of $27.46$ GHz ESR field at rotating \Cupz\ sample in the $bc$ plane; $T=1.3$ K. Solid
  symbols correspond to actual experimental data, and open symbols
  are the repetition of solid symbols with the 180$^\circ$ period. Symbols size corresponds to
  resonance intensity. Dashed lines are guide to an eye.  A rough sketch of a two-domain sample and orientation of
  $x$,$y$,$z$
axes within domains is shown.}\label{FIG:rosettes}
\end{figure}

\begin{figure}
  \includegraphics[width=0.45\textwidth]{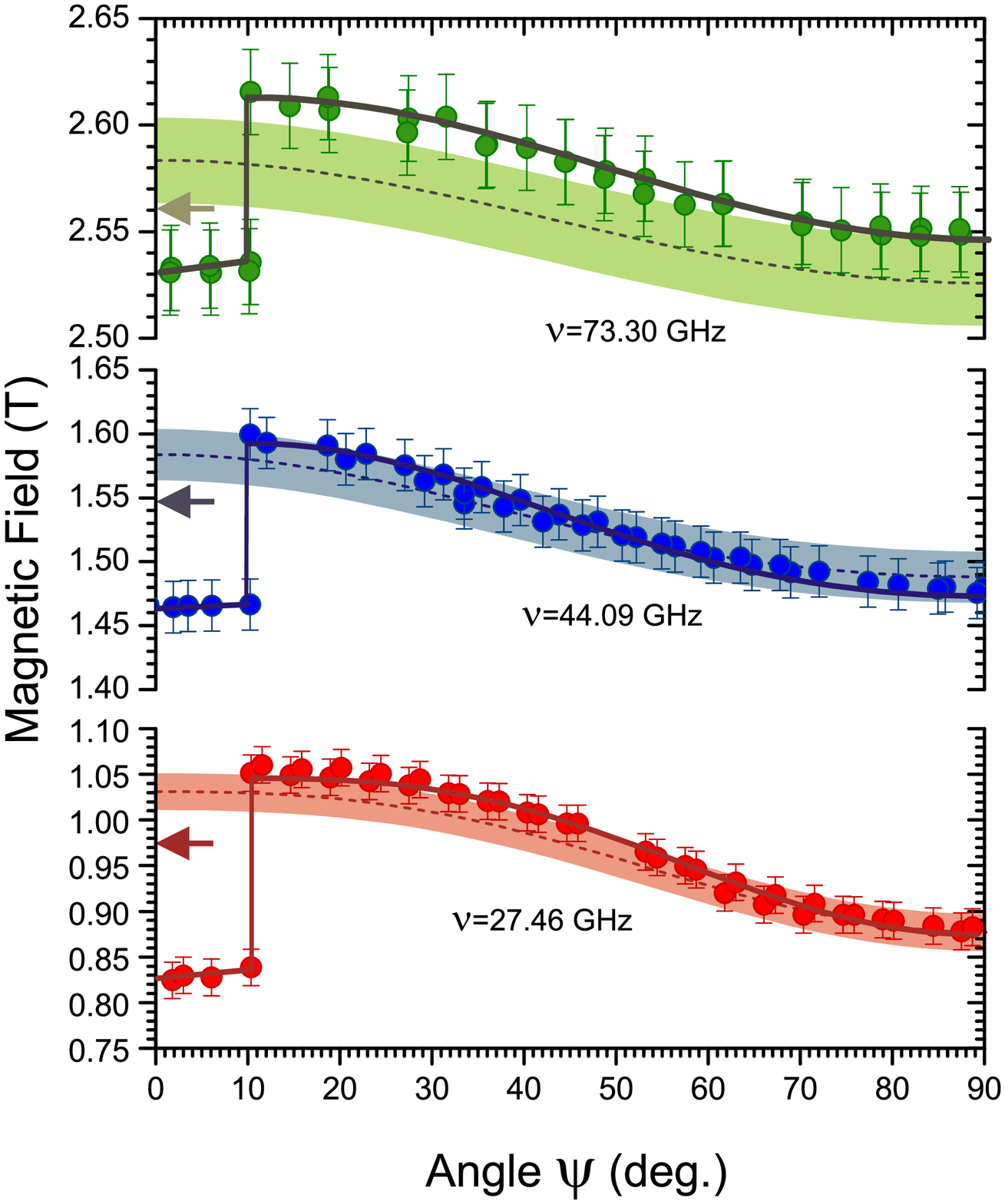}\\
  \caption{(Color online) Summary of angular dependencies in $xy$-plane for a domain. Solid lines are guide for the eye,
  dashed lines are theoretical calculation (biaxial model) with a shaded region around, marking a possible error due to the parameters uncertainty.
  Arrows indicate the field of paramagnetic resonance at corresponding frequency.}\label{FIG:AnomalyAngularvarFQ}
\end{figure}

\begin{figure}
  \includegraphics[width=0.45\textwidth]{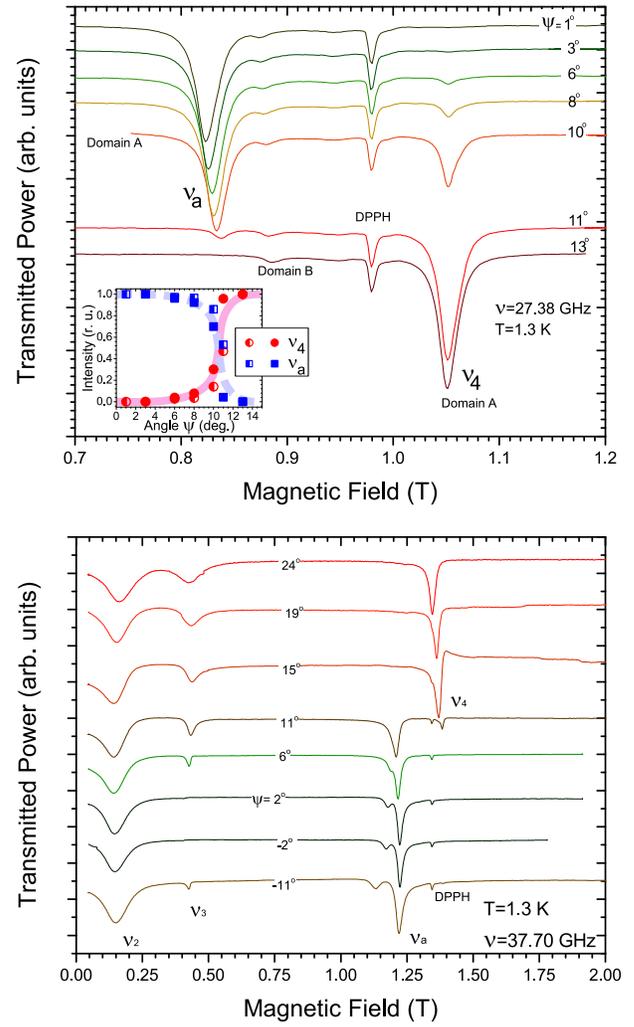}\\
  \caption{(Color online).
  Evolution of the ESR line of a single domain sample at the rotation of the magnetic field within the $xy$-plane
  near the critical angle $\psi_c=10^\circ$, when regular mode $\nu_{4}$ transforms to anomalous mode $\nu_{a}$.
  Zero field cooling from 15 K to $T=1.3$ K was performed before each record.  Upper panel: $\nu=27.28$ GHz.
  Lower panel: $\nu=37.7$ GHz.
  Inset: Angular dependences of the intensities of ESR modes $\nu_{4}$ (red circles) and $\nu_{a}$ (blue squares).
  Each half-open symbol corresponds to a line recorded after zero-field cooling.
  Solid symbols present ESR lines recorded without thermocycling, each rotation performed
  in a field 1.2 T. Thick lines are guide for an eye.}\label{FIG:27GHz-slow}
\end{figure}

Magnetic resonance signal in \Cupz\ at temperatures above $T_{N}=4.2$ K corresponds to a typical
exchange-narrowed paramagnetic resonance of Cu$^{2+}$ ions with anisotropic $g$--factor. The values of
$g$--factor, obtained by high-temperature ($T\gtrsim10$ K) ESR measurements  are $g_{x}=g_{y}=2.05$ and
$g_{z}=2.28$. A narrow Lorentzian  line, with a halfwidth of about $5\cdot10^{-3}$ T, broadens with cooling, and
becomes unresolvable near $T_{N}$. Below $T_{N}$ the ESR response becomes strongly anisotropic.  For ${\bf H}
\parallel z$ a broad signal transforms into a single narrow line, shifted from high-temperature position, while
for ${\bf H}\parallel x,y$ two lines appear, as shown on left panel
of Fig. \ref{FIG:heating}. The resonance halfwidth shows a clear
critical dependence near the phase transition temperature. The
divergency in the line halfwidth together with the shift of the
resonance position, as shown on Fig. \ref{FIG:TevolPoints}, can be
used as a marker of phase transition, allowing us to extract $T_{N}$
from the ESR data. The value of $T_N$=4.2$\pm0.1$ K is in agreement
with the results of magnetization measurements, as shown on the
phase diagram (Fig. \ref{FIG:PhasdiagrJoint}).

\subsection{Antiferromagnetic resonance  \label{SECTIONResultsESR_Domain}}

\begin{figure}
    \includegraphics[width=0.45\textwidth]{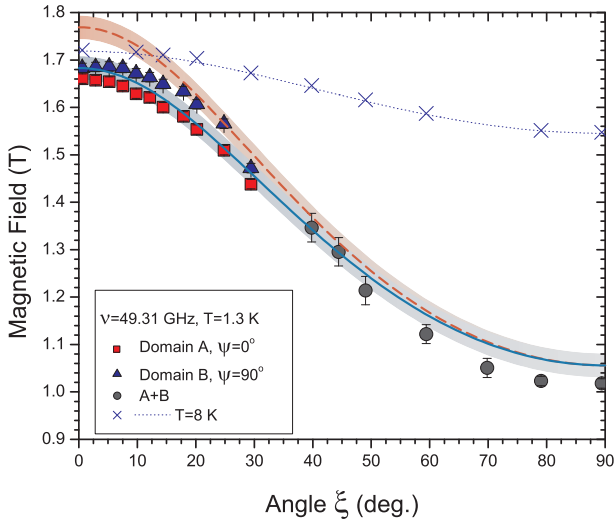}\\
  \caption{(Color online) Angular dependence of the field of $49.31$ GHz ESR in
  a two-domain sample of \Cupz\ at the temperature $T=1.3$ K.
Rotation is performed in the xz-plane of the domain A. Triangles
--- domain A,
  squares -- domain B, circles -- signal of the whole sample when domains become indistinguishable, crosses
  --
  $T=8$ K ). Solid line presents the calculated low-temperature resonance field (biaxial model)for $\psi=90^\circ$,
  dashed line - for $\psi=0$ and dotted line - theory for high-temperature paramagnetic resonance.
  Shaded region marks error
  boundaries of model calculation (see text).}\label{FIG:RotTheta49GHz}
\end{figure}

\begin{figure}
  \includegraphics[width=0.45\textwidth]{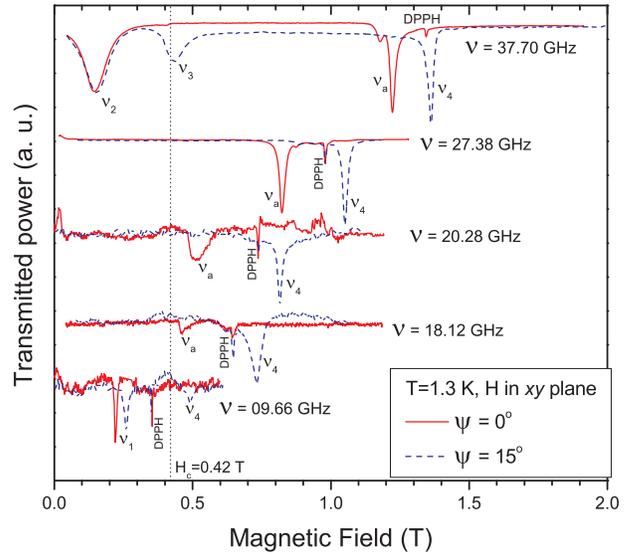}\\
  \caption{(Color online) Collection of ESR lines at $T=1.3$ K for fields along $\psi=0^{\circ}$ (red solid) and $\psi=15^{\circ}$ (blue dashed). Vertical
  dashed line denotes critical field $H_{c}=0.42$ T at this temperature.}\label{FIG:CurvesCollection}
\end{figure}

 The anisotropy within the $bc$ plane below $T_N$ results in the angular dependence of the
resonance field, as  shown on Figs. \ref{FIG:rosettes}, \ref{FIG:AnomalyAngularvarFQ}. This reveals  two kinds
of resonances with the identical rosette-like  angular dependencies, which are shifted for 90$^\circ$ on Fig.
\ref{FIG:rosettes}. The relation between the intensities of these two kinds of signals is different for
different samples.  This observation indicates a presence of two kinds of domains. The ratio of intensity of
signals from two kinds of domains has the same value for zero-field cooling and field-cooling of the sample in
the field of 6 T, as well as at thermocycling through $T_N$. Therefore we conclude, that these domains are
crystallographic domains, for which $b$ axes are rotated for 90$^\circ$. We denote domains with orthogonal $b$
($c$) axes as domain A and domain B. One of the samples has the intensity for one rosette much stronger than for
another one. For this, approximately single domain sample, the relation between the volumes of domains of
different types may be evaluated, e.g., from the relation of intensities of ESR lines presented on
Fig.\ref{FIG:27GHz-slow}, upper panel, between 0.82 and 0.88 T. Such an estimation gives the number of spins,
belonging to domain A, approximately 30 times greater than to domain B. This sample remained approximately a
single domain one at numerous cycles of cooling from the room temperature.

The rosettes shown on Fig.\ref{FIG:rosettes} demonstrate a smooth
evolution of the resonance field with the angle in the whole angle
range except for the narrow range in the vicinity of $b$-direction.
This direction was identified for the nearly single domain sample by
room temperature X-ray diffraction. At the angle $\psi=10^\circ$
there is a step-like jump of the resonance field, shown in
Fig.\ref{FIG:rosettes} and Fig. \ref{FIG:AnomalyAngularvarFQ}. Near
the exact orientation of the external field along $b$-axis, i.e.
when tilting $\psi$  does not exceed 10$^\circ$, the resonance field
is shifted to much lower field and this position can not be
extrapolated from the smooth angular dependence in the main part of
the field range. Therefore we denote the resonance observed at
$\mid\psi\mid<10^\circ$ as anomalous mode $\nu_a$. The
redistribution of the intensity from the regular to the anomalous
mode at a slow rotation of the field is shown on Fig.
\ref{FIG:27GHz-slow}. One can see here that the transmission of the
intensity between the two types of resonances has a character of a
switching, it is performed within an interval of about 1$^\circ$
which may be a measure of the mosaic of the sample. Thus, a narrow
phase transition at the angle variation is observed.

\begin{figure}
  \includegraphics[width=0.5\textwidth]{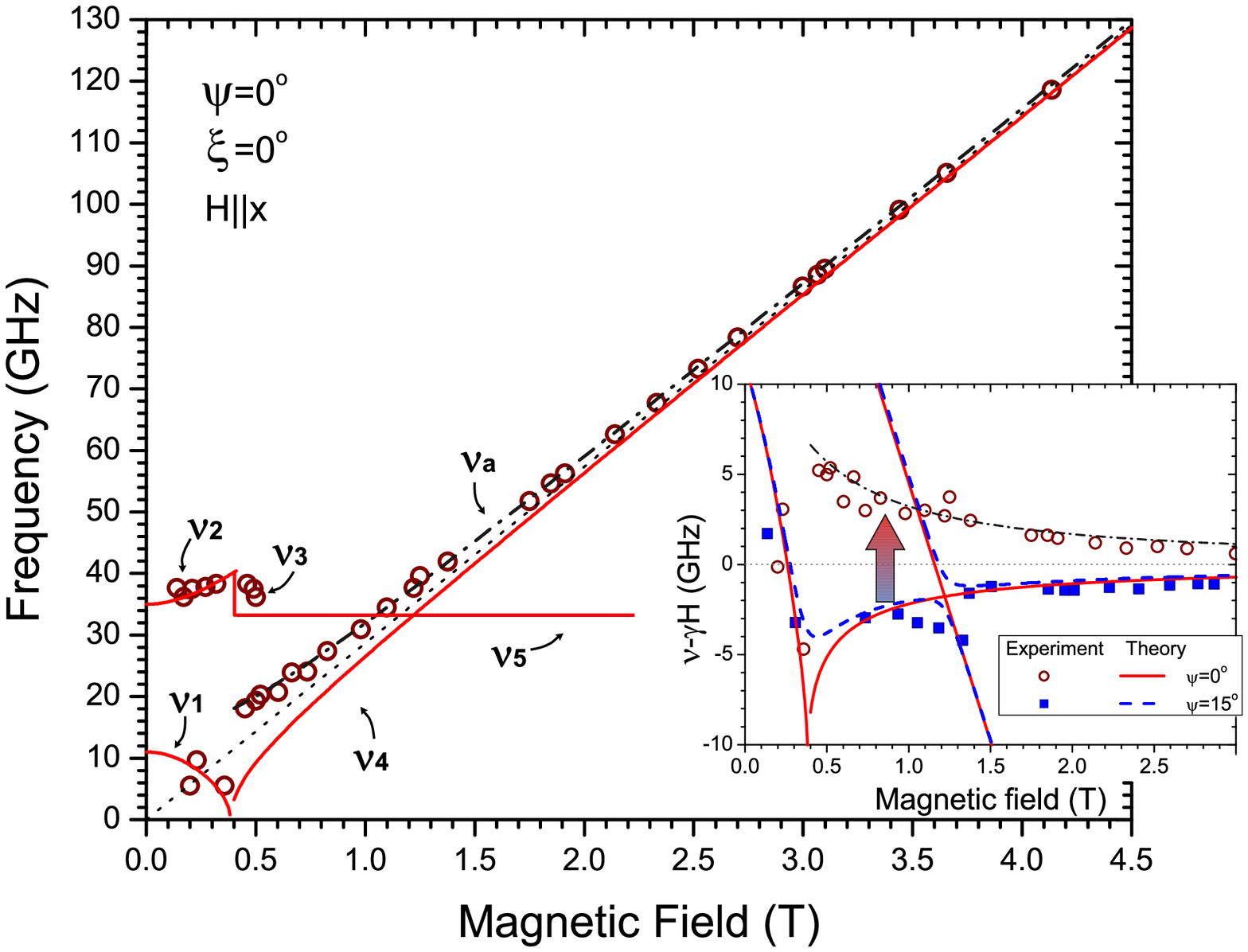}\\
  \caption{(Color online) AFMR spectra at $T=1.3$ K for field along easy axis. Solid line is the model calculation, dash-dotted line is
  the empirical formula (\ref{anomode}). Insert: shift from paramagnetic resonance frequency vs. field for directions $\psi=0^{\circ}$ (empty circles)
  and $\psi=15^{\circ}$ (solid squares) in $xy$ plane. Solid and dashed lines are theoretical calculations, corresponding to these cases.
  Dash-dotted line is the same, as on main plot.}\label{FIG:SpectraXY0and15}
\end{figure}

\begin{figure}
  \includegraphics[width=0.45\textwidth]{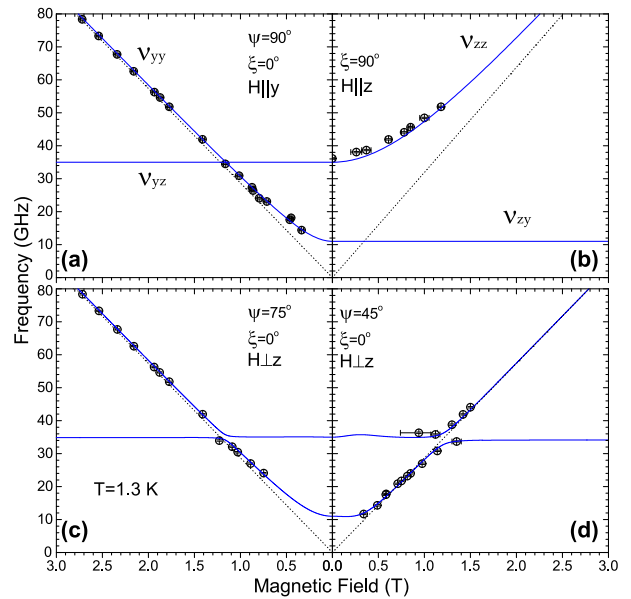}\\
  \caption{(Color online) AFMR spectra for several field directions in $xy$ plane and along $z$ axis at $T=1.3$ K. Solid lines are calculation according
  to (\ref{Lagr_general}) with $\Delta_{y}=11$ GHz and $\Delta_{z}=35$ GHz,
  dashed lines are paramagnetic resonance with $g_{xy}=2.05$ and $g_{z}=2.28$.}\label{FIG:SpectraXY45-75-90-Z}
\end{figure}

We didn't observe any difference between the resonance signals,
indicating this transition, when passing the critical angle in a
field or in zero-field, and for passing the critical angle at the
temperature above or below $T_N$. Tilting the field within the
$xz$-plane  conserves the anomalous mode, as shown on Fig.
\ref{FIG:RotTheta49GHz}, at least in the range $\mid\xi\mid\lesssim
30{^\circ}$, where the difference between the anomalous mode and an
extrapolation for a regular mode may be detectable. The anomalous
mode was observed only at $\mu_0 H>0.4$ T. Below this field the
resonance positions at $\psi=0^\circ$ and $\psi=15^\circ$ are almost
identical, as one can see on upper and lower records of
Fig.\ref{FIG:CurvesCollection} and on the low-field part of
frequency-field dependencies on the inset of Fig.
\ref{FIG:SpectraXY0and15}. At the same time, at $\mu_0 H>0.4$ T
there is a jump-like evolution of the resonance field and frequency
in this range of angles.

   Further, we measured ESR fields for a set of frequencies (see examples of records on Fig.\ref{FIG:CurvesCollection})
at three principal directions of the magnetic field and at a tilting angle $\psi=15^\circ$, as well as for two
intermediate orientations in $bc$ plane. The corresponding frequency-field dependencies are presented on Figs.
\ref{FIG:SpectraXY0and15},\ref{FIG:SpectraXY45-75-90-Z}. From these data we conclude, that the spectrum of
frequencies of the antiferromagnetic resonance has two energy gaps, approximately equal to 35 and 10 GHz and two
branches in a magnetic field. For the direction of the magnetic field near $b$-axis there is a mode softening at
approaching the field of 0.42 T from the zero-field side. At $\mu_0 H>0.42$ T we observe the softened mode in
the angular range of the regular mode and the anomalous mode in the narrow angle range $\mid \psi \mid
<10^\circ$. By changing the angle $\psi$ across the critical value toward $\psi=0$, the ESR frequency is
transposed from the value below the paramagnetic resonance frequency $g_x\mu_B \mu_0 H/2\pi \hbar$ to the value
above it. This transposition is marked by an arrow on the inset of Fig. \ref{FIG:SpectraXY0and15}, it occurs by
a jump at crossing the critical angle $\psi_c=10^\circ$. This jump corresponds exactly to the jump of the
resonance field, shown on Fig. \ref{FIG:AnomalyAngularvarFQ}.

 We note here that the observed frequency-field dependencies for field orientation in the whole solid angle, except
 for the range of the anomalous mode, may be well described by the calculated frequencies of the two sublattice
 antiferromagnet with a biaxial anisotropy and the easy axis  directed along $b$,  see, e.g. Ref.\onlinecite{NagamyaAF}.
 The calculated frequencies are given in the Appendix A and presented
 on Figs. \ref{FIG:SpectraXY0and15}, \ref{FIG:SpectraXY45-75-90-Z} by solid lines. Eight curves shown here, and
 the calculated angular dependencies shown on Fig. \ref{FIG:RotTheta49GHz}
 are parameterized only by two energy gaps and three $g$-factors. The $g$-factors $g_x$, $g_y$, $g_z$ are measured
 independently in the paramagnetic phase and are not fitting parameters.  Below the
 critical field of 0.42 T, the frequencies in the whole solid angle range of the magnetic field directions are described
 with that model.
  In particular, a mode softening  at ${\bf{H}}\parallel b$
 indicates the spin-flop transition. By this observation we
 can conclude that $b$ is the easy axis direction.  For  the anomalous mode, observed at $\mid\psi\mid<10^\circ$,
 $\mid\xi\mid<30^\circ$, $\mu_0 H>0.42$ T
we use the empirical relation

\begin{equation}
    \nu_{a}=\sqrt{\Delta_{a}^{2}+(\frac{g_x\mu_{B}}{2\pi\hbar}\mu_0 H)^{2}}
    \label{anomode}
\end{equation}

with $\Delta_{a}=14$ GHz at $T=1.3$ K.  This relation represents the
observed frequency at the unexpected position above (and not below)
the paramagnetic resonance frequency at $H>H_c$.

Thus, the ESR data reveal a weak magnetic anisotropy in $bc$ plane
and a spin-flop transition, as well as the anomalous mode $\nu_a$
appearing in the narrow angular range of the field direction instead
of a regular resonance of a biaxial antiferromagnet.

\section{Magnetization \label{SECTIONResultsPPMS}}
\subsection{Field along the easy axis \label{SECTIONResultsPPMS_Hx}}

\begin{figure}
    \includegraphics[width=0.5\textwidth]{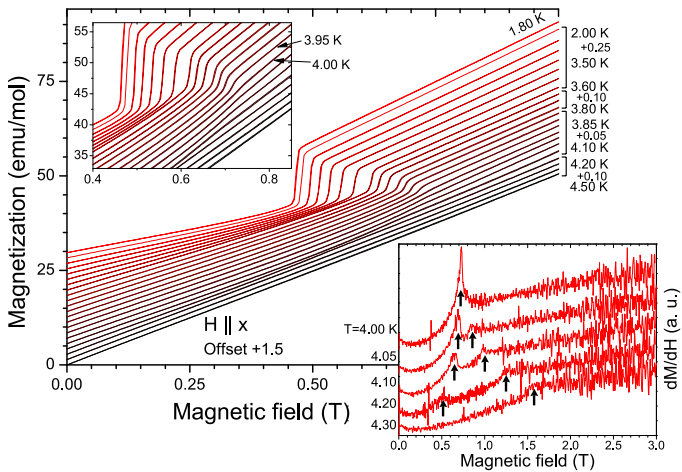}\\
  \caption{(Color online) Isothermal low-field magnetization curves of \Cupz, field along $x$.
  An offset of $1.5$ emu/mol per curve is present. Upper insert: an
  expanded region around the temperature, where magnetization jump
  disappears.
  Lower insert: a
  few $dM/dH$ curves are shown,
  allowing to locate phase transitions between disordered and ordered phases, marked by arrows.}\label{FIG:XmhCollection}
\end{figure}

\begin{figure}
    \includegraphics[width=0.5\textwidth]{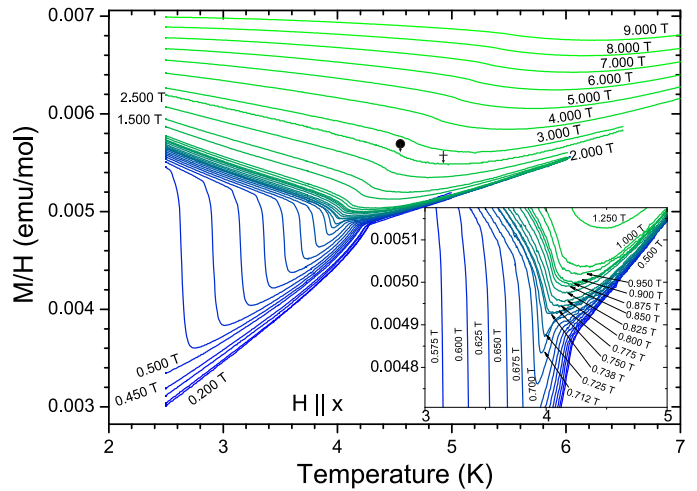}\\
  \caption{(Color online) Normalized magnetization $M(T)/H$ for various magnetic fields, directed along $x$.
  Insert: an expanded region around the point, where magnetization jump vanishes.}\label{FIG:XmtCollection}
\end{figure}

The main feature of  low-temperature magnetization curves at ${\bf H} \parallel b$  is the presence of jump in
magnetization corresponding to the spin-flop transition, detected by ESR. As it is shown at Fig.
\ref{FIG:XmhCollection}, magnitude of the jump increases with cooling, and its position shifts to lower fields.
Jump in magnetization disappears around $4\pm0.05$ K, which is lower than $T_{N}$. The sharp increase in
magnetization can also be seen in $M(T)$ curves, present on Fig. \ref{FIG:XmtCollection}. Crossing the spin-flop
phase boundary by temperature in a constant field also gives a very pronounced step in magnetic moment. This
step disappears at $\mu_0 H=0.74$ T, and above this field there appears a minimum on the
 $M$ $vs$ $T$ dependence. Below the temperature of the minimum of the magnetization there is a kink,
 marking the onset of long range order. The minimum and the kink are marked on Fig. \ref{FIG:XmtCollection}.
 Both minimum and
kink shift upwards in temperature with increasing field up to 9 T,
though the former becomes less pronounced. Note, that there is  no
offset on Fig. \ref{FIG:XmtCollection}, and stacking of the curves
reflects the nonlinearity of magnetization process.

\begin{figure}
    \includegraphics[width=0.5\textwidth]{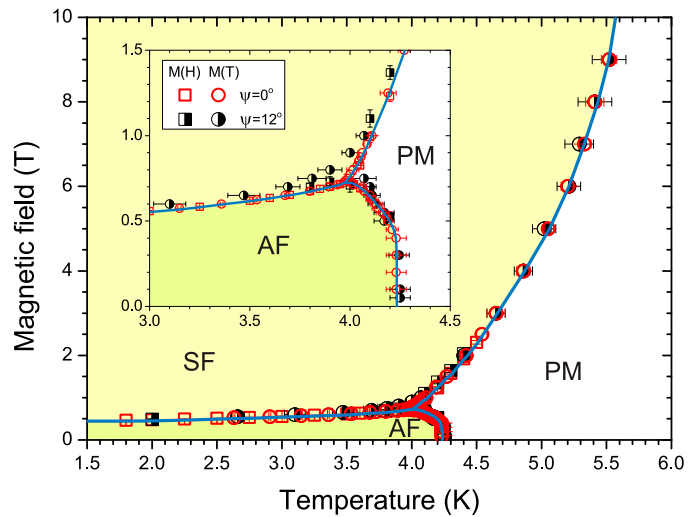}\\
  \caption{(Color online) Phase diagram of \Cupz, field along easy axis. AF is $\mathbf{l}\parallel x$ collinear antiferromagnetic phase,
  SF is spin-flop antiferromagnetic phase and PM
  is paramagnetic phase. Circles are features in $M(T)$ curves, squares are features in $M(H)$; red points correspond
  to $\psi=0^{\circ}$ orientation, and black --- to $\psi=12^{\circ}$. Lines are guide for an eye. In the insert an
  expanded region around the bicritical point is shown.}\label{FIG:psi00phasediagram}
\end{figure}

We derive the ordering point by a peak in the derivative $\partial (MT)/\partial T$, as suggested by
Fisher\cite{Fisher}. We can also detect this transition by a peak in the derivative $dM/dH$ of isothermal
magnetization curve, as displayed at the insert of Fig. \ref{FIG:psi00phasediagram}. A final phase diagram, with
points on phase boundaries obtained by both $M(T)$ and $M(H)$ scans, is present at Fig.
\ref{FIG:psi00phasediagram}.

\subsection{Field in $bc$ plane \label{SECTIONResultsPPMS_Hxy}}

\begin{figure}
    \includegraphics[width=0.5\textwidth]{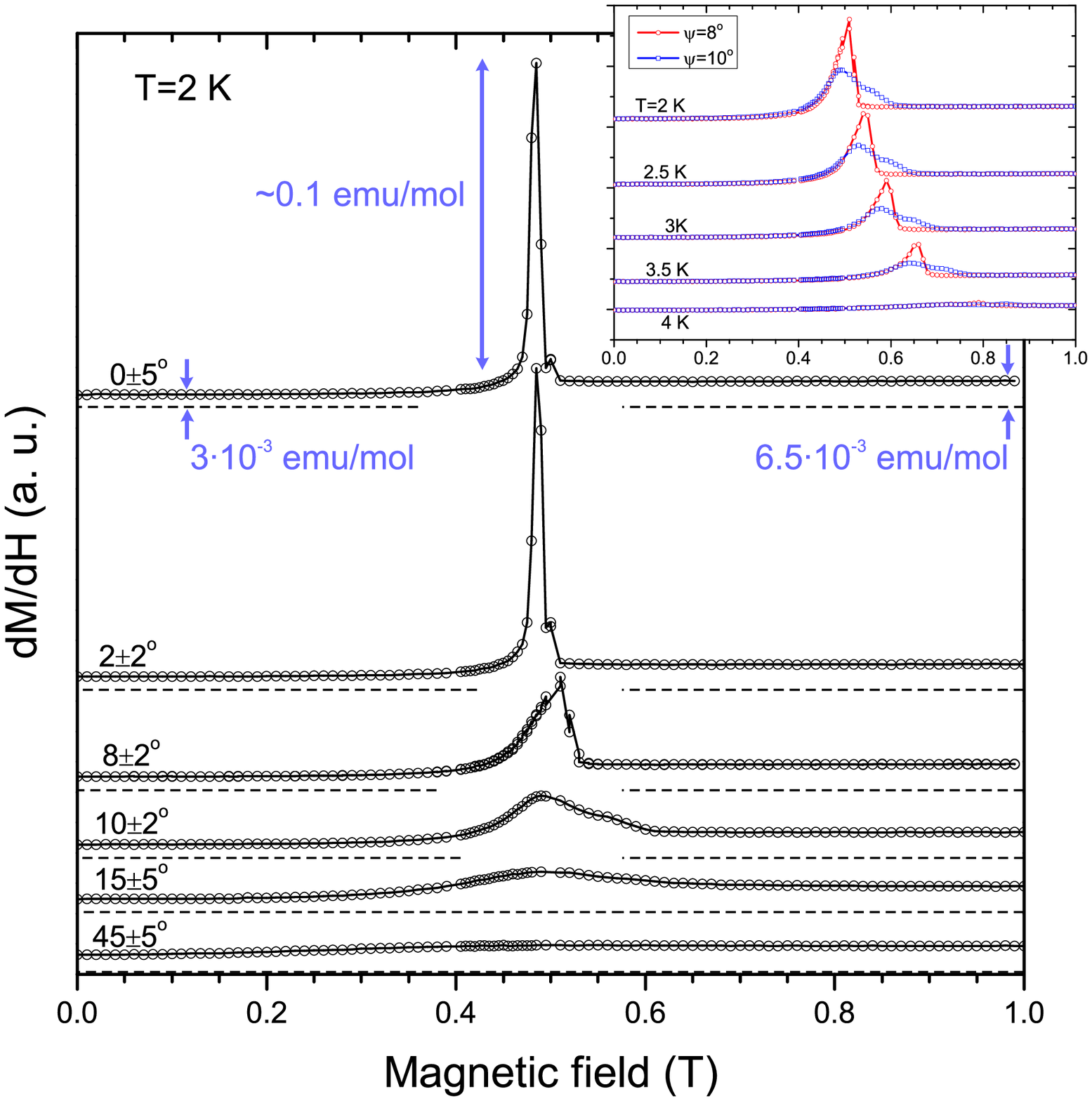}\\
  \caption{(Color online) Magnetization derivative $dM/dH$ for various directions of magnetic field in $xy$ plane at $T=2$
  K. Dashed lines show zero for corresponding curves.
  Inset: temperature dependencies of magnetization derivatives for directions $\psi=8^{\circ}\pm2^{\circ}$
  and $\psi=10^{\circ}\pm2^{\circ}$.}\label{FIG:mhDerivatives}
\end{figure}

On Fig. \ref{FIG:mhDerivatives} a collection of magnetization derivatives $dM(H)/dH$ for various directions of
magnetic field in $xy$ plane is present. $dM(H)/dH$ curves show a sharp peak at $\psi\rightarrow0$, which
broadens with misalignment. Left wing of the peak is always smooth, in contrast to a discontinuity in the right
wing of the peak. The discontinuity exists up to a critical angle $\psi_c$, observed in angular dependences of
ESR, see Figs. \ref{FIG:rosettes}, \ref{FIG:AnomalyAngularvarFQ}. When tilt angle exceeds $\psi_{c}$, the
$dM/dH$ curve becomes completely smooth. The transition between smooth and discontinuous types of derivative,
which affects mostly the right wing, occurs abruptly. This difference between the curves, corresponding to field
tilts below and above $\psi_{c}$, is pronounced in a whole temperature range where spin-flop transition takes
place, as shown in the inset of Fig. \ref{FIG:mhDerivatives}. The discontinuity of differential susceptibility
$dM/dH$ occurs exactly at the same magnitude and in the same angular range $\mid\psi\mid<\psi_c$, as the
anomalous ESR mode $\nu_a$.

With further increase in $\psi$ peak becomes less pronounced and
almost disappears when $\psi$ approaches $\sim45^{\circ}$.
Magnetization curve at $\sim45^{\circ}$ (this is the direction along
natural crystal facets) doesn't show a step, and demonstrates a
smooth slope increase, as observed in Ref. \onlinecite{XYbehavior}.

We have also performed a study of a phase diagram with magnetic field, slightly tilted from $x$ axis. The
orientation we chose was $\psi=12\pm2^{\circ}>\psi_{c}$. Here \Cupz\ still demonstrates increase in
magnetization near $H_{c}$, but the transition is regular, i.e. with a smooth derivative $dM/dH$ on both sides
of $H_c$. We locate $H_{c}$ and $T_{N}$ in the same way as was described before for ${\bf H} \parallel x$. The
resulting phase boundaries are also shown at Fig. \ref{FIG:psi00phasediagram}, and the difference between phase
diagram for exact and misaligned orientations along $x$ is observed only near the bicritical point.

\subsection{Field along middle and hard axes \label{SECTIONResultsPPMS_Hyz}}

\begin{figure}
    \includegraphics[width=0.5\textwidth]{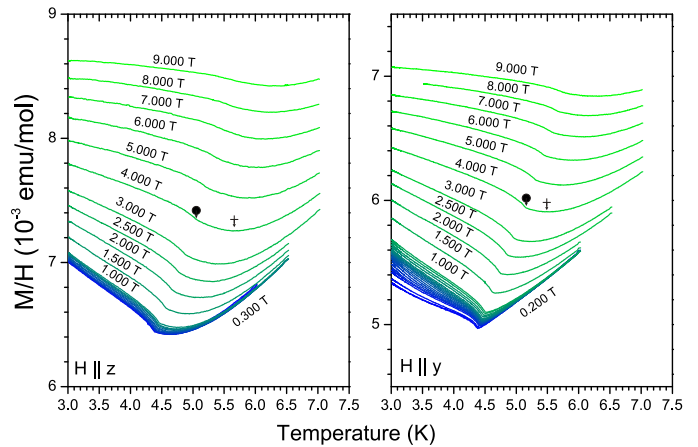}\\
  \caption{(Color online) Normalized magnetization $M(T)/H$ for various magnetic fields, directed along $z$ (left panel) and $y$ (right panel).
  Values of magnetic field between $0.3$ and $1$ Tesla are the same, that on Fig. \ref{FIG:XmtCollection}. The kink and
  the minimum on the curves are marked by a droplet and a cross, respectively.}
  \label{FIG:YZmtCollection}
\end{figure}

Curves of normalized magnetization for the fields, directed along middle axis $y$ and hard axis $z$ at Fig.
\ref{FIG:YZmtCollection} don't show spin-flop, but demonstrate an increase of $T_{N}$ in a magnetic field. The
only qualitative difference between this two sets of curves is in the onset of minimum of $M(T)$: for field
along $y$ the minimum appears for the applied fields  between   $1.5$ and $2$ T, while for field along $z$ it is
present even at $H\rightarrow0$ (as independently confirmed by zero-field ac magnetization
measurements\cite{XYbehavior}). Another feature of $M(T)$ curves at ${\bf H}
\parallel y$ is the existence of inflection point below $T_{N}$ at low fields, in contrast to the case of ${\bf
H} \parallel z$. As the field is increased, the inflection disappears.

\section{Discussion \label{SECTIONDiscussion}}
\subsection{Biaxial model and spectra \label{SECTIONDiscussion_Biaxial}}

In section \ref{SECTIONResultsESR} we have presented AFMR spectra for various field directions. These spectra
follow biaxial collinear antiferromagnet paradigm for all magnetic field directions, except for a small solid
angle corresponding to anomalous mode. The anomalous mode is observed in  a solid angle of about of
$10^{-2}4\pi$, close to easy axis, and only above $H_{c}$. Resonance frequencies in this range of fields and
angles correspond to a gapped branch (\ref{anomode}) with $\Delta_{a}=14$ GHz. This unexpected effect can be
described as an in-plane anisotropy switching caused by spin flop, i.e. the resonant frequencies are that of a
two sublattice biaxial antiferromagnet, for which $x$ turns abruptly from the easy- into middle axis and $y$
turns into easy axis at the spin-flop point. This conclusion is made on the base of the experimental observation
of the anomalous mode, which appears at $H>H_c$ and has the frequency following the relation (A3), corresponding
to middle axis orientation of the field, instead of expected (A7), derived for the easy axis orientation.

 We consider the
magnetoelastic hypothesis, which might explain the switching of
anisotropy at the spin flop point. The in-plane anisotropy, marking
the easy axis, originates from rhombic distortion of square lattice,
for \Cupz\ this distortion is due to a relative difference of about
$\sim10^{-4}$ between the lattice constants $b$ and $c$. In
principle, the antiferromagnetic ordering may cause a striction of
the same order of magnitude\cite{MorosinStriction}. Because of the
magnetostriction, the in-plane anisotropy may be dependent on the
magnitude and the direction of sublattice magnetizations, and,
therefore, it should change when the spin flop transition takes
place. One could expect the $y$ axis to become easy axis, and $x$ to
become middle axis immediately after the spin-flop.

Nonetheless, a simple quantitative formulation of this approach, described in details in Appendix
\ref{SECTIONMagnetoelastic}, does not capture a step-like angular dependence of AFMR field in $xy$-plane, and is
in a contradiction with the observed relation between the zero-field gap and a critical field. The analysis of a
complete Lagrangian,  allowed by symmetry, should include several dozens of magnetoelastic and elastic terms and
was not performed.

Another possibility, presumably explaining the nature of anomalous mode $\nu_{a}$, is the existence of a phase,
other than collinear for $H>H_{c},\psi<\psi_{c}$. This implies destabilizing of collinear phase by frustration
when external field compensates in-plane anisotropy. However, our measurements do not support this hypothesis:
magnetization curve $M(H)$ and phase boundary $T_{N}(H)$ are indistinguishable for $\psi<\psi_{c}$ and
$\psi>\psi_{c}$ in fields above $H_{c}$.

The influence of a change of the direction and magnitude of zero point fluctuations at the spin flop may be also
of importance, because anisotropic spin fluctuations also contribute to the energy of anisotropy.

Nevertheless, the nature of the anisotropy switching remains unclear.

To give a connection with the previous work [\onlinecite{neutrons}]
we derive a relation between zero-field gaps $\Delta_{y,z}$ of
antiferromagnetic resonance and the parameters of microscopic model
Hamiltonian. In nearest-neighbor exchange approximation the complete
biaxial Hamiltonian reads as

\begin{equation}
\begin{aligned}
    &\hamilt=\sum\limits_{\aver{i,i'}}J\spop_{i}\spop_{i'}-g\mu_{B} \mu_0 \sum\limits_{i}\mathbf{H}\spop_{i}\\
    &\phantom{HHH}-\sum\limits_{\aver{i,i'}}\left(\delta J_{y}\spin^{y}_{i}\spin^{y}_{i'}+\delta
    J_{z}\spin^{z}_{i}\spin^{z}_{i'}\right),
    \label{hamiltonian}
\end{aligned}
\end{equation}
 where $\delta J_{y}$ and $\delta J_{z}$ are parameters of so-called   'exchange anisotropy'.
 According to linear spin-wave approximation, which have proven
 to be good for describing the $k$-dependence of the spectrum in the vicinity of Brillouin
 zone center of \Cupz\ [\onlinecite{domain}], the energy gaps are
 related to exchange anisotropy parameters as

\begin{equation}\label{EQgapsrefinement}
     \Delta_{y,z}=2\sqrt{2J\delta J_{y,z}}.
\end{equation}

In this spin-wave approximation the sublattice magnetization is
supposed to be $\mu_B$ per magnetic ion, which is not the case of
the \Cupz, where a strong quantum reduction of about 50 $\%$ is
observed. A finer estimation for the case of $S=1/2$ SLAFM,
considering $1/S$ corrections, was given by Weihong \textit{et
al}\cite{WeihongGap}:

\begin{equation}\label{EQgapsrefinement2}
     \Delta_{y,z}\simeq1.2\sqrt{2J\delta J_{y,z}}.
\end{equation}

Thus, this equation may be used for an estimation of $\delta
J_{y,z}$. Spectroscopic gaps of $35\pm2$ and $11\pm2$ GHz are
$\Delta_{z}=1.68\pm0.1$ K and $\Delta_{y}=0.53\pm0.1$ K
correspondingly. Hence, from (\ref{EQgapsrefinement2}) we extract
$\delta J_{z}=53.2$ mK and $\delta J_{y}=5.3$ mK. This corresponds
to relative exchange anisotropy $\delta J_{z}/J=3.1\cdot10^{-3}$ and
$\delta J_{y}/J=3.1\cdot10^{-4}$. This is in agreement with previous
neutron data, except for the parameter $\delta J_{y}$, which was not
resolved by neutron scattering experiment. We can characterize the
observed anisotropy switching, in terms of changing of parameters of
Hamiltonian (\ref{hamiltonian}). It corresponds to transformation of
$\delta J_{y}$ into $\delta J_{y}^{\ast}$, which is of
\emph{negative} sign and equals $-6.7$ mK.

\subsection{Phase diagrams  \label{SECTIONDiscussion_Phases}}

\begin{figure}
    \includegraphics[width=0.5\textwidth]{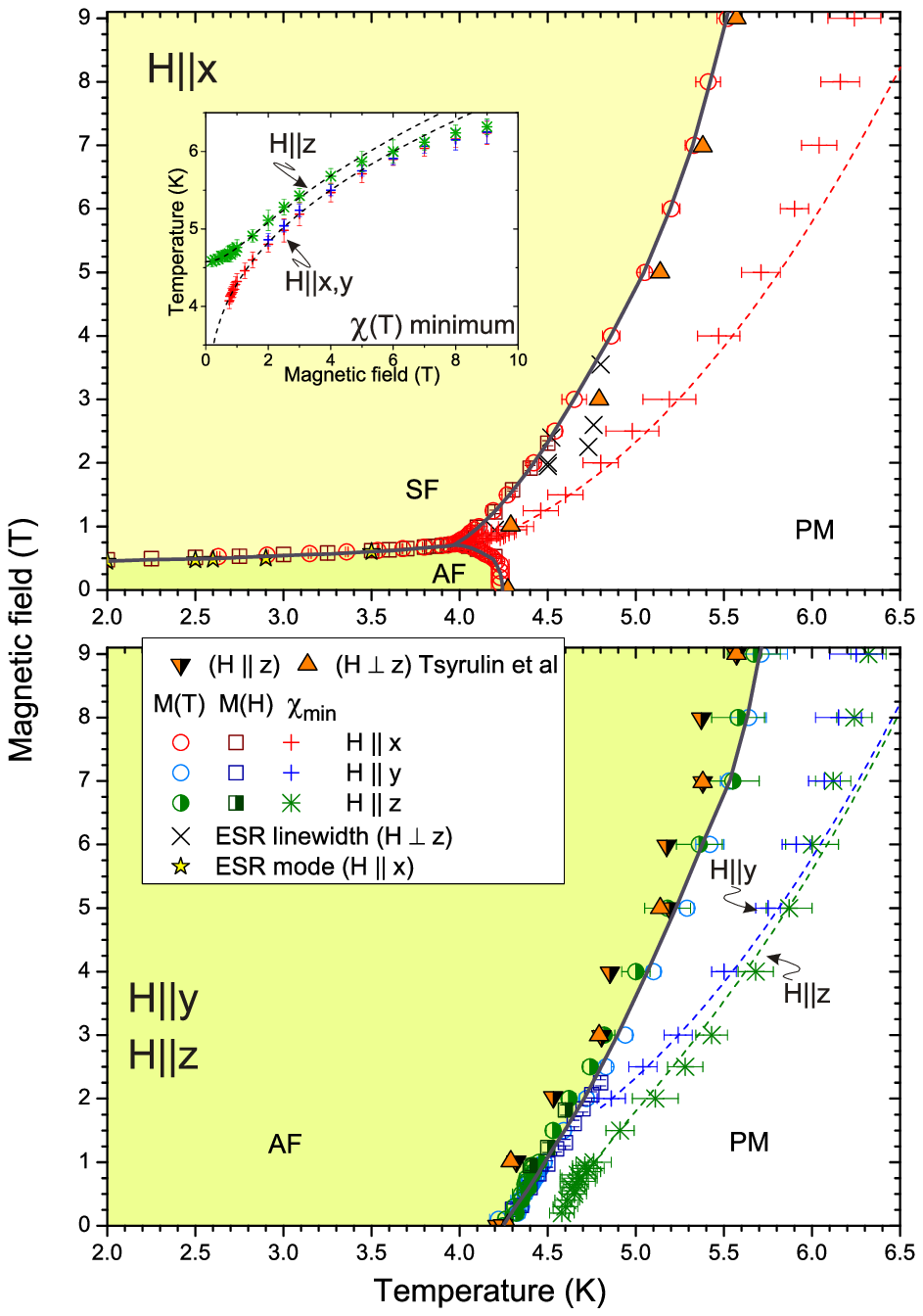}\\
  \caption{(Color online) Phase diagrams for $H\parallel x$ (upper panel) and $H\parallel yz$ (lower panel) directions. Points are experimental data: circles for $M(T)$ and squares for $M(H)$ features.
  Stars and diagonal crosses mark phase transitions determined by ESR, solid triangles are data from [\onlinecite{domain}].
  Horizontal crosses ($x$, $y$) and snowflakes ($z$) are minima in $M(T)$. Solid lines are guide for the eye, dashed lines are fit (\ref{EQTminfit}) for $T_{min}(H)$
  The fit for all three orientations is separately shown in the insert in the upper panel.}\label{FIG:PhasdiagrJoint}
\end{figure}

Phase diagrams for $x$, $y$ and $z$ directions of magnetic field are
present at Fig. \ref{FIG:PhasdiagrJoint}. For $y$ and $z$ directions
the phase diagram are analogous, with monotonous increase of
$T_{N}$. For field along $x$ phase diagram is more complicated, with
a bicritical point, where spin-flop, ordered and paramagnetic phases
meet. The  phase diagram  presents the spin flop transition and the
bicritical point in addition to the phase boundaries reported in the
previous work using neutron scattering and specific heat
measurements\cite{domain}.

The field, at which the antiferromagnetic resonance mode  $\nu_{3}$  is observed, also marks the spin-flop
transition (see Appendix). From  ESR experiment we get $\mu_0 H_{c}^{ESR}=0.45$ T at $T=1.3$ K (see Fig.
\ref{FIG:SpectraXY0and15}), this field increases with temperature. Temperature dependence of $H_{c}$ derived
from ESR is consistent with the magnetization measurements as shown on Fig. \ref{FIG:PhasdiagrJoint}.

 Minima of $M(T)$ are also plotted, showing different behavior for all three directions.
While for ${\bf H} \parallel z$ minimum persists up to $H=0$ limit, for different directions it appears only at
some finite field, at which $T_{min}$ reaches $T_{N}$.

The reason of this minimum may be qualitatively explained by the following consideration: for an easy-pane
antiferromagnet it is natural to have an anisotropic susceptibility, which is larger for out-of-plane direction.
In case of 2D antiferromagnet with $\delta J_{z}\ll J$ one should expect this anisotropic behavior to rise only
at low temperatures, when $T<J$. Numerical simulations of HSLAF with a weak easy-plane
anisotropy\cite{XYcucolliPRL} show, that tendency for increase of magnetization at  $\bf{H}||z$ due to the onset
of planar correlations overcomes the tendency for its decrease due to short-range AF order. Thus a
characteristic  minimum in $\chi(T)$ marks a crossover from Heisenberg to $XY$ behavior. Cuccolli \textit{et al}
\cite{XYcucolliPRL}, using a QMC data analysis, suggested a formula for estimation of $T_{min}$ in a case of
easy-plane HSLAF model,

\begin{equation}\label{EQTminfit}
    T_{min}=\dfrac{4\pi\rho_{s}}{\ln\left(\frac{C}{\delta_{eff}}\right)}.
\end{equation}

Here $\rho_{s}\simeq0.22J$ is the renormalized spin stiffness,
$\delta_{eff}$ is the relative anisotropy and $C\simeq160$ is the
dimensionless constant. It has also been found, that presence of
external magnetic field in 2D magnets makes them effectively
easy-plane and induces both Berezinsky-Kosterlitz-Thouless
transition at finite temperature and a minimum in $\chi(T)$ above it
[\onlinecite{CucolliField}]. In absence of long-range order, when a
\emph{local} order parameter $\mathbf{l}$ is formed, the orientation
$\mathbf{l}\perp\mathbf{H}$ provides an energy gain. Hence, the
short-range order parameter becomes 2D instead of 3D in the
isotropic case and the effective anisotropy energy in this
field-induced $XY$ behaviour is proportional to $H^{2}$ .

For the orientation of the magnetic field $\mathbf{H}\perp z$ for a
strong enough field ($g\mu_{B}\mu_0 H\gtrsim\sqrt{\delta J_{z}J}$),
we consider 'effective' easy-plane anisotropy induced by an external
field. The easy plane of this anisotropy is perpendicular to the
field. We take this anisotropy in the form derived in Ref.
\onlinecite{CucolliField} for HSLAFM

\begin{equation}\label{EQdeltaeffXY}
    \delta_{eff}^{xy}=\beta\left(\dfrac{g_{xy}\mu_{B}\mu_0 H}{k_{B}J}\right)^{2},
\end{equation}

where $\beta$ is a dimensionless parameter. Here we disregard smaller anisotropy $\delta J_{y}$, as the
experimental $T_{min}(H)$ in $x$ and $y$ directions is the same within the error bars.

For the case $\mathbf{H}\parallel z$ the easy-plane anisotropy
originates due to the combination of the natural and field-induced
anisotropy. We empirically combine these two factors which were
analyzed separately in QMC simulations\cite{CucolliField,CucolliXXZ}

\begin{equation}\label{EQdeltaeffXY}
    \delta_{eff}^{z}=\dfrac{\delta
    J_{z}}{J}+\beta\left(\dfrac{g_{z}\mu_{B}\mu_0 H}{k_{B}J}\right)^{2}.
\end{equation}

Fitting experimental data for $T_{min}$ with equation
(\ref{EQTminfit}), where $\delta_{eff}$ is set as
$\delta^{xy}_{eff}$, or $\delta^{z}_{eff}$, and parameters $J=18.1$
K and $C=160$ are fixed, we yield $\rho_{s}\simeq0.24J$, $\delta
J_{z}\simeq0.023$ K  and $\beta\simeq0.1$, which is quite close to
the result of Cucolli \textit{et al} [\onlinecite{CucolliField}].
Fits are shown on Fig. \ref{FIG:PhasdiagrJoint} with dashed lines.
This result can be considered as another indication of the 2D
correlations developing in \Cupz\ at $T>T_N$. Nonetheless, value of
$\delta J_{z}$ obtained by this fit is in better agreement with
estimation by Eq. \ref{EQgapsrefinement}, which does not take into
account quantum renormalization of the gap, than with Eq.
\ref{EQgapsrefinement2}, which considers $1/S$ corrections.



For ${\bf H} \parallel x$ (i.e. field along the easy axis)  there is a bicritical point $(H_{c},T_{c})$. Three
phase transition lines meet in this point:  second order paramagnetic to collinear antiferromagnetic phase
transition (PM-AF), second order paramagnetic to flopped antiferromagnetic phase transition (PM-SF) and first
order spin-flop phase transition AF-SF. In the vicinity of bicritical point the following scaling equations are
expected\cite{KostNelFish}: for AF-SF transition temperature dependence for critical field is

\begin{equation}\label{EQscalingAFSF}
    H^{2}(T)-H_{c}^{2}=A\left(\frac{T}{T_{c}}-1\right),
\end{equation}

while for ordering transitions to AF and SF phases relations between
magnetic field and ordering temperatures are

\begin{equation}\label{EQscalingPMAF}
    H^{2}(T)-H_{c}^{2}=A\left(\frac{T}{T_{c}}-1\right)-B_{AF}\left(\frac{T}{T_{c}}-1\right)^{\phi}
\end{equation}

and

\begin{equation}\label{EQscalingPMSF}
    H^{2}(T)-H_{c}^{2}=A\left(\frac{T}{T_{c}}-1\right)+B_{SF}\left(\frac{T}{T_{c}}-1\right)^{\phi}
\end{equation}

correspondingly. For 'classical' 3D antiferromagnet scaling exponent is known to be $\phi=1.25$ for uniaxial and
$\phi=1.175$ for biaxial anisotropy. Theory also suggests amplitude ratio $Q=B_{SF}/B_{AF}=1$ for the former
case\cite{KostNelFish, RohrerKingMnF2}. In contrast, for pure 2D case with easy-axis anisotropy bicritical point
is expected\cite{NelsonPelkovits2D,Transition2D,WeirdRb2MnF4} to occur only at $T=0$ --- a simple argument for
that is the following: when easy-axis anisotropy is compensated by the external field, the system becomes
equivalent to non-perturbed two-dimensional Heisenberg model, which can possess long-range order only at zero
temperature. The PM-AF and PM-SF phase boundaries, which meet at $T=0$, are defined by

\begin{equation}\label{EQscalingExponentional}
    |H^{2}(T)-H_{c}^{2}|\propto T^{-2}\exp\left(-\frac{4\pi\rho_{s}}{T}\right).
\end{equation}

\begin{figure}
    \includegraphics[width=0.5\textwidth]{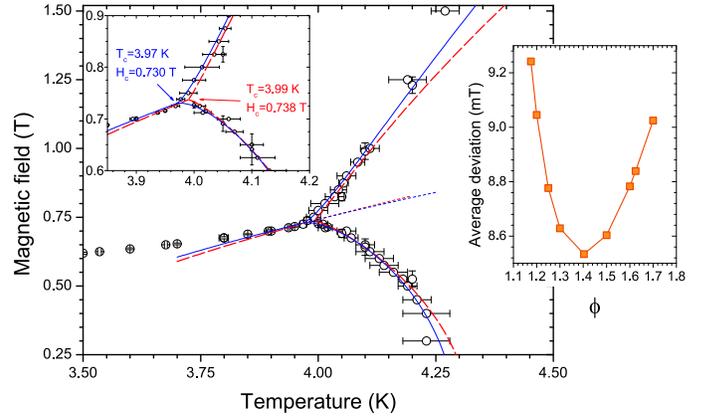}\\
  \caption{(Color online) Scaling in the vicinity of bicritical point. Blue solid and red dashed lines are fits (\ref{EQscalingAFSF},\ref{EQscalingPMAF},\ref{EQscalingPMSF}) with
  $\phi=1.4$ and $\phi=1.175$ correspondingly, dotted lines are extension of (\ref{EQscalingAFSF}) to the $T>T_{c}$ region.
  Left insert: magnified part of main plot. Right insert: fit deviation
  (see text) versus critical exponent $\phi$.}\label{FIG:Bicritical}
\end{figure}

The above equation is valid only in the absence of additional
anisotropies and interlayer couplings, while in case of \Cupz\ both
of these perturbations are present and the bicritical point is at
$T>0$. A numerical proof for the latter statement can be found,
e.g., in Monte-Carlo study of classical anisotropic $XY$
antiferromagnet on square lattice
[\onlinecite{Deutsche_phasediagram}], where the phase diagram
strongly resembles that of a 3D easy-axis AFM. Hence, phase diagram
of \Cupz\ turns out to be an intermediate case between ideal 2D and
conventional 3D anisotropic antiferromagnets. Straightforward fit of
experimental data with equations
(\ref{EQscalingAFSF},\ref{EQscalingPMAF},\ref{EQscalingPMSF}) gives
bicritical point at $T_{c}=3.97$ K, $\mu_0 H_{c}=0.73$ T with
scaling exponent $\phi=1.4$ and amplitude ratio $Q=1.78$. Fixing the
value of $\phi$ to theoretically suggested for 3D antiferromagnet
$\phi_{3D}=1.175$ leads to $T_{c}=3.99$ K, $\mu_0 H_{c}= 0.738$ T
and $Q=1.54$, but with a worse fit quality. Data and fits are
present at Fig. \ref{FIG:Bicritical}, together with numerical
quality criterion
--- sum of average least squares for all three formulas
(\ref{EQscalingAFSF},\ref{EQscalingPMAF},\ref{EQscalingPMSF}). It can be concluded that reliable estimation of
universal parameters from our data is $\phi=1.4\pm0.2$ and $Q=1.8\pm0.2$, and bicritical point is located at
$\mu_0 H_{c}=0.730\pm0.006$ T and $T_{c}=3.97\pm0.03$ K. Region $\Delta T\simeq0.2$ K where the scaling
equations are fulfilled is about 5\% of  $T_{c}$, which is significantly larger than for classical
three-dimensional uniaxial antiferromagnet MnF$_{2}$ ($\Delta T/T_{c}\sim10^{-3}$
[\onlinecite{RohrerKingMnF2}]), though smaller than for quasi-2D compound Rb$_{2}$MnF$_{4}$ ($\Delta
T/T_{c}\sim0.26$ [\onlinecite{WeirdRb2MnF4}]), with a purely uniaxial anisotropy. These facts, as well the
larger value of critical index $\phi=1.4>\phi_{3D}$ result in a conclusion, that \Cupz\ presents an intermediate
behaviour between 3D and 2D models in the vicinity of bicritical point.

The observed dependence of the phase diagram in the bicritical point range  on the field orientation  is
natural, because bicritical point is very sensitive to field misalignment\cite{RohrerKingMnF2}. On the contrary,
away from the region around $H_{c}$,  the phase boundaries for $\psi=0^{\circ}$ and $\psi=12^{\circ}$ coincide
within error bars.

\section{Conclusions \label{SECTIONConclusions}}

In the present work we have studied AFMR spectra and magnetization
curves of HSLAFM \Cupz. These measurements reveal the presence of
biaxial anisotropy in \Cupz, instead of easy-plane formulation used
earlier. From the ESR experiments we have derived two energy gaps
$\Delta_{z}\simeq 35$ and $\Delta_{y}\simeq 11$ GHz. The weak
in-plane anisotropy is responsible for the spin-flop phase
transition at $\mu_0 H_{c}\simeq0.4$ T in $\mathbf{H}\parallel x$
direction. The AFMR spectra also show, that weak in-plane anisotropy
is changing its sign at the spin-flop transition. This anisotropy
reversal, occuring  in a manner of switching,  may be also observed
as a phase transition at changing the orientation of the magnetic
field within the $bc$ plane, at the critical angle 10$^\circ$ with
respect to the easy axis direction. The conjecture that this anomaly
might be of magnetoelastic origin may, by simplified treatment,
explain the anisotropy reversal at the spin flop, but is not
consistent with the step-like angular dependence of ESR field or
frequency. It is also not consistent  with the observed relation
between the energy gap $\Delta_y$ and spin-flop field $H_c$.  The
nature of the abrupt reversal of the weak anisotropy remains
unclear.

The hypothesis which is probably worth to analyze theoretically is a possible change of the direction and
magnitude of zero point spin fluctuations at the spin flop. A change of the contribution of fluctuations to the
energy of the anisotropy may also change the effective anisotropy of the ordered spin component.

The field-dependence of the temperature of the minimum on $M(T)$
curves for three principal orientations are found to be in agreement
with the results of numerical simulation of HSLAFM
\cite{XYcucolliPRL}. The increase of $T_{N}$ in external field has
been found for all orientations in agreement with previous
measurements. Scaling exponent $\phi=1.4\pm0.2$ of phase boundaries
near the bicritical point is intermediate between 2D and 3D models.

Accounting for the observed weak anisotropy might be significant for correct estimation of other weak
interactions, e.g.  next-nearest neighbor and interlayer exchange from experimental data\cite{J1J2newtheory}.
Similar anisotropy can be present in another HSLAFM's of Cu-pz family (namely, Cu(pz)$_{2}$(BF$_{4}$)$_{2}$ and
[Cu(pz)$_{2}$(NO$_{3}$)](PF$_{6}$)), as according to Xiao's magnetization data\cite{XYbehavior} there are
signatures of spin-flop transitions as well.

We note that for deuterated \Cupz\ the difference between $b$ and $c$, resulting in the weak anisotropy, is
larger than for a regular sample with hydrogen, so it would be of interest to test if the in-plane anisotropy is
the same in a deuterated sample. Other possible future experiments include NMR and neutron scattering with field
along $x$ to probe the magnetic structure as well as search for anomaly in magnetic field dependencies of
low-temperature elastic properties (magnetostriction, ultrasound propagation etc).

\section{Acknowledgments}
The authors would like to thank O. S. Volkova and A. N. Vasiliev
(Moscow State University), and A. Zheludev (ETH Z\"{u}rich) for the
opportunity of using their experimental facilities and assistance
with it. Also special thanks to V. N. Glazkov, L. E. Svistov, S. S.
Sosin, V. I. Marchenko, M. E. Zhitomirsky and W. E. A. Lorenz for
fruitful and stimulating discussions. This work was supported by
RFBR Grant No 12-02-00557.

\vspace{20mm}

\appendix
\section{AFMR frequencies of a biaxial antiferromagnet\label{SECTIONBiaxial}}

 A theory for AFMR in a two-sublattice antiferromagnet with
biaxial anisotropy has been developed in 1950's\cite{NagamyaAF}. For both orientations of the magnetic field
along hard- and middle-axis ($z$ and $y$), in the ground state we have the antiferromagnetic order parameter
$\mathbf{l}\parallel x$, and this orientation of $\mathbf{l}$ is independent of external field magnitude.

Magnetic resonance frequencies for $\mathbf{H}
\parallel z$ are
\begin{align}
    \label{Wz_1} &\nu_{zz}=\sqrt{(\frac{g_z\mu_{B}}{2\pi\hbar}\mu_0 H)^{2}+\Delta_{z}^{2}},\\
    \label{Wz_2} &\nu_{zy}=\Delta_{y};
\end{align}

and for $\mathbf{H} \parallel y$ we have
\begin{align}
    \label{Wy_1} &\nu_{yy}=\sqrt{(\frac{g_y\mu_{B}}{2\pi\hbar} \mu_0 H)^{2}+\Delta_{y}^{2}},\\
    \label{Wy_2} &\nu_{yz}=\Delta_{z}.
\end{align}

At the orientation of the magnetic field along the easy axis the case is more complicated, as the ground state
is field-dependent. There is a spin-flop transition with an abrupt change from  $\mathbf{l}\parallel x$ to
$\mathbf{l}\parallel y$. The critical field of this transition is
\begin{equation}
    \mu_0 H_{c}=2\pi\frac{\hbar\Delta_{y}}{g_x\mu_{B}}
    \label{spinflop}
\end{equation}
This transition is accompanied by a jump in magnetization. The ESR frequencies below and above $H_c$ are the
following:

For $H<H_{c}$:

\begin{widetext}
\begin{equation}
\label{Wx_12}\nu_{1,2}=\sqrt{(\frac{g_x\mu_{B}}{2\pi\hbar}\mu_0
H)^{2}+\frac{\Delta_{y}^{2}+\Delta_{z}^{2}}{2}\mp\sqrt{2(\frac{g_x
\mu_{B}}{2\pi\hbar}\mu_0 H)^{2}(\Delta_{y}^{2}+\Delta_{z}^{2})+(\frac{\Delta_{y}^{2}-\Delta_{z}^{2}}{2})^{2}}}\\
\end{equation}
\end{widetext}

 For $H>H_{c}$
\begin{align}
    \label{Wx_2-} \nu_{4}&=\sqrt{(\frac{g_x\mu_{B}}{2\pi\hbar}\mu_0 H)^{2}-\Delta_{y}^{2}}\\
    \label{Wx_2+} \nu_{5}&=\sqrt{\Delta_{z}^{2}-\Delta_{y}^{2}}
\end{align}

The resonant mode $\nu_{3}$ with the vertical  $\nu_{3}(H)$ dependence corresponds to spin-flop transition, as
the system is allowed to absorb energy in a band of frequencies in the critical point. Modes $\nu_{1}$ and
$\nu_{4}$ are softened at the critical field $H_{c}$. For the intermediate field orientations we calculated the
frequencies of spin resonance numerically within the same formalism.

\section{Magnetoelastic correction\label{SECTIONMagnetoelastic}}
For description of ESR modes at $T\rightarrow0$  we use  macroscopic exchange symmetry
formalism\cite{AndrMarcUFNeng}. This formalism, in particular, reproduces the results of a  mean-field theory of
a two-sublattice antiferromagnet with biaxial anisotropy\cite{NagamyaAF}. In the framework of the exchange
approach the spin structure is considered to be collinear, and the anisotropy of a relativistic origin, and
magnetization,induced by the external field, are taken as perturbations. Though being applicable only in fields
$H\ll H_{sat}$, this formalism is model-independent and allows easy introduction of additional anisotropy terms.
As the saturation field in \Cupz\ constitutes almost 50 T, restriction on the field magnitude is not an issue
for the exchange symmetry formalism applicability. Our calculations are based on the following Lagrange function
per mole of the compound (in CGS system):

\begin{equation}\label{Lagr_veryGeneral}
    \mathcal{L}=\frac{\chi_{\perp}}{2\gamma^{2}}\left(\dot{\mathbf{l}}+
{\gamma}[\mathbf{H}\times \mathbf{l}]\right)^{2}-U_{a}.
\end{equation}

Here $\gamma=\dfrac{g\mu_{B}}{\hbar}$ is the gyromagnetic ratio,
unit vector  $\mathbf{l}$ with the orientation, given by angles
$\varphi$ and $\theta$ shown on Fig.\ref{FIG:rosettes}, is the order
parameter, $\mathbf{H}$ is magnetic field; $\chi_{\perp}$ is the
magnetic susceptibility in the direction, perpendicular to
$\mathbf{l}$. Equation (\ref{Lagr_veryGeneral}) also implies
$\chi_{\parallel}=0$ at zero temperature. Term $U_{a}=\eta
l_{y}^{2}+\zeta l_{z}^2$ is the anisotropy energy. We assume
positive constants $\eta$ and $\zeta$, $\eta<\zeta$. Hence
$\mathbf{l}\parallel x$ minimizes anisotropy energy.

The ground state and magnetic resonance frequencies may be
calculated using this Lagrange function, as described in
Ref.\onlinecite{AndrMarcUFNeng}. The ground state and the spectrum
are identical to that of Ref. \onlinecite{NagamyaAF}, described
above. The anisotropy constants may be expressed via energy gaps:
$\eta=\chi_{\perp}(2\pi\Delta^{2}_{y})/2\gamma^{2}$ and
$\zeta=\chi_{\perp}(2\pi\Delta^{2})_{z}/2\gamma^{2}$, where
$\Delta_{z}>\Delta_{y}$ are the energy gaps, which one actually
observes in the ESR experiment. With this substitution, Lagrange
function (\ref{Lagr_veryGeneral}) is

\begin{equation}
\begin{split}
\label{Lagr_general} \mathcal{L}=
\frac{\chi_{\perp}}{\gamma^{2}}\left(\frac{1}{2}\left(\dot{\mathbf{l}}+
{\gamma}[\mathbf{H}\times \mathbf{l}]\right)^{2}-{}\right. \phantom{abcdefghijk}&\\
-
\left.\frac{(2\pi\Delta_{z})^{2}}{2}l_{z}^{2}-\frac{(2\pi\Delta_{y})^{2}}{2}l_{y}^{2}\right)&,
\end{split}
\end{equation}

and corresponding potential energy in non-zero magnetic field is

\begin{equation}
\begin{split}
\label{Epot_general} \mathcal{E}=
\frac{\chi_{\perp}}{\gamma^{2}}\left(-\frac{\gamma^{2}}{2} [\mathbf{H}\times \mathbf{l}]^{2}+{}\right. \phantom{abcdefghijklmnop}&\\
+
\left.\frac{(2\pi\Delta_{z})^{2}}{2}l_{z}^{2}+\frac{(2\pi\Delta_{y})^{2}}{2}l_{y}^{2}\right)&,
\end{split}
\end{equation}

\begin{figure}
    \includegraphics[width=0.5\textwidth]{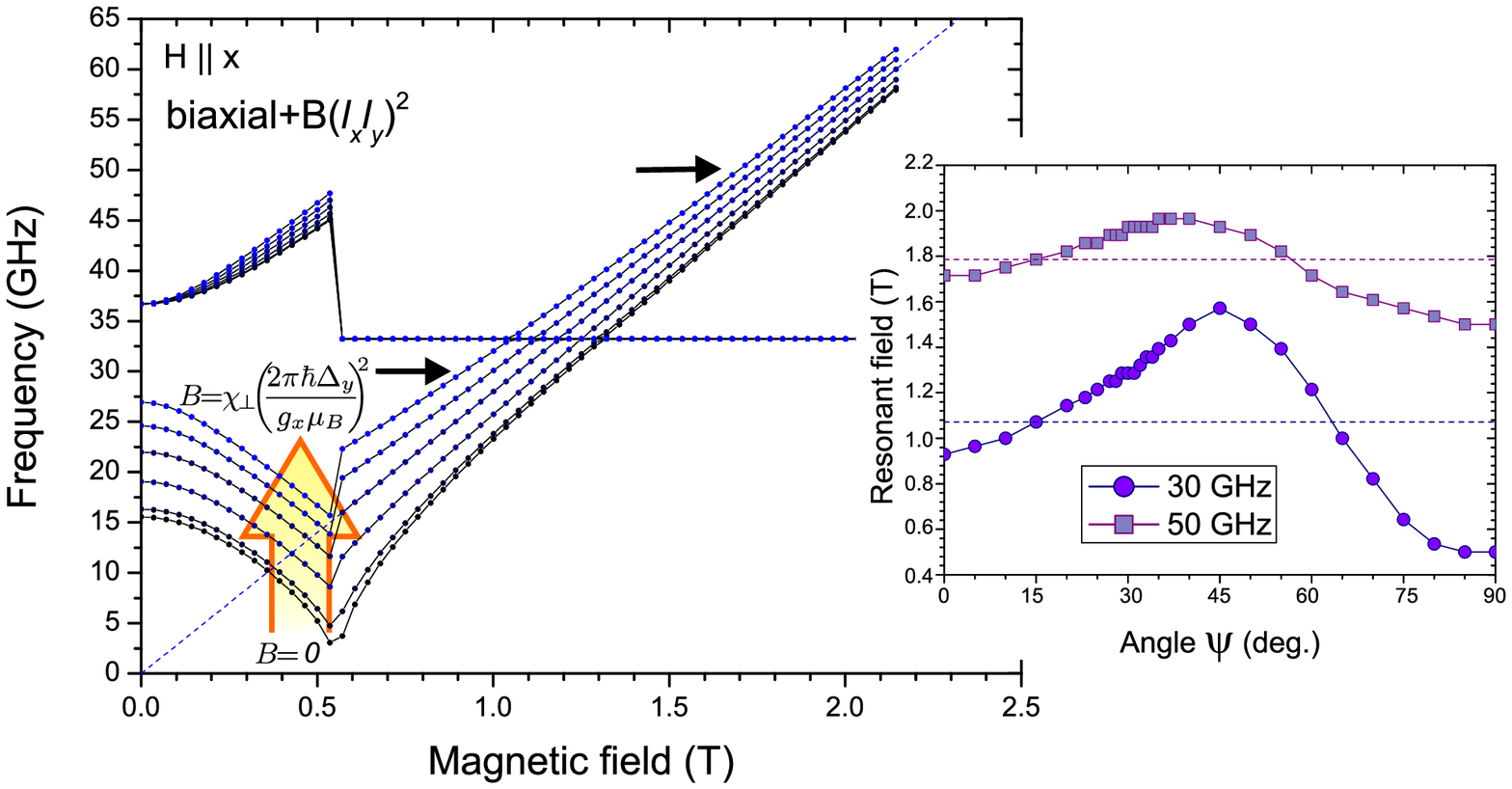}\\
  \caption{(Color online) Spectra of model (\ref{Lagr_quartic1}) for various values of quartic term. Field is directed along $x$. Arrows mark frequencies
  at which angular dependencies for $B=\chi_{\perp}\left(\dfrac{2\pi\hbar\Delta_{y}}{g_{x}\mu_{B}}\right)^{2}$ are shown in the insertion;
  dashed lines are the expectation for paramagnetic resonance.}\label{FIG:Quartic1Joint}
\end{figure}

\begin{figure}
    \includegraphics[width=0.5\textwidth]{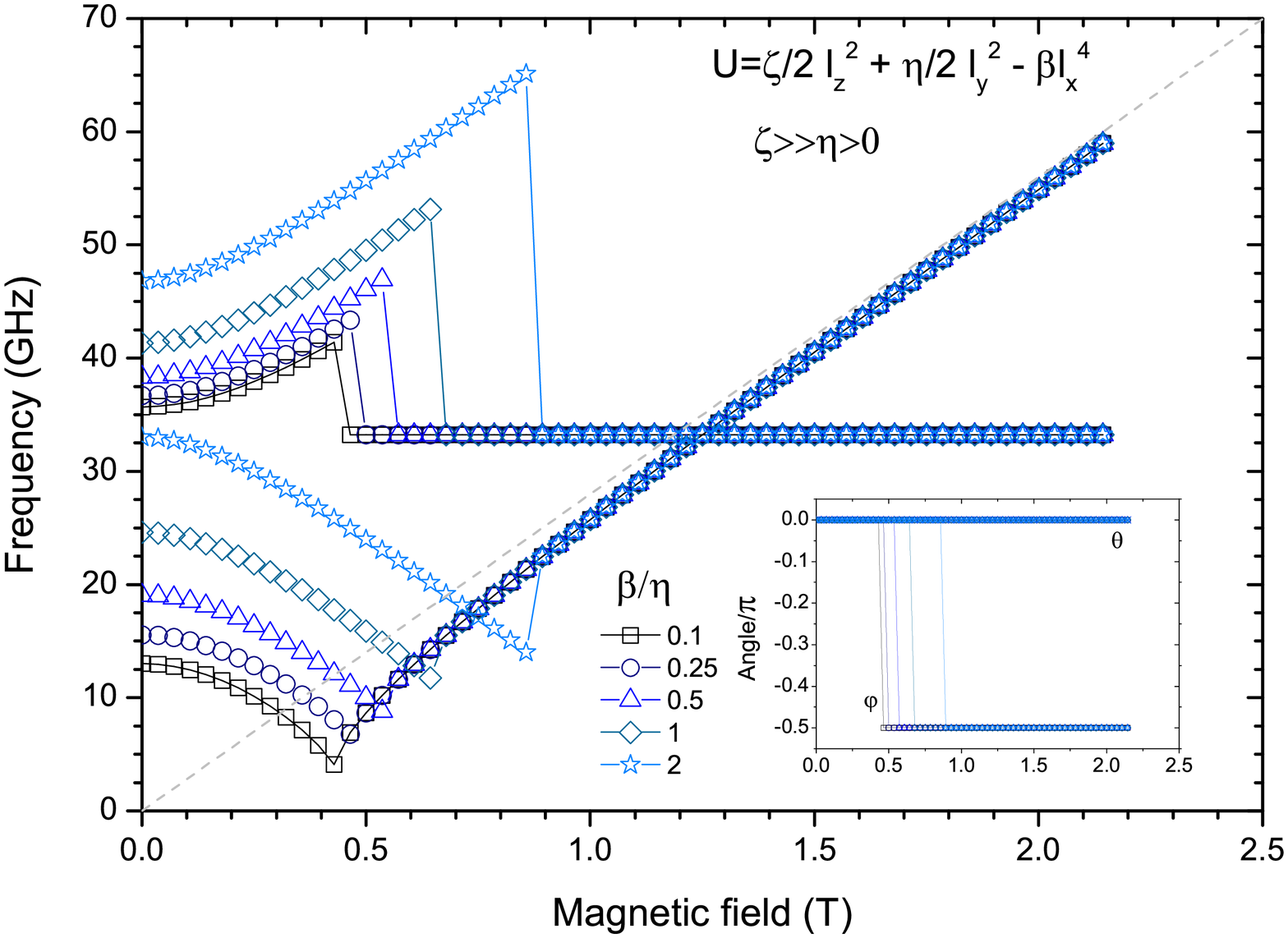}\\
  \caption{(Color online) Spectra of the model with a quartic term $-\beta l_x^4$, ${\bf H}\parallel x$.
  Dashed line is paramagnetic resonance.} \label{FIG:plusbetax}
\end{figure}
\begin{figure}
    \includegraphics[width=0.5\textwidth]{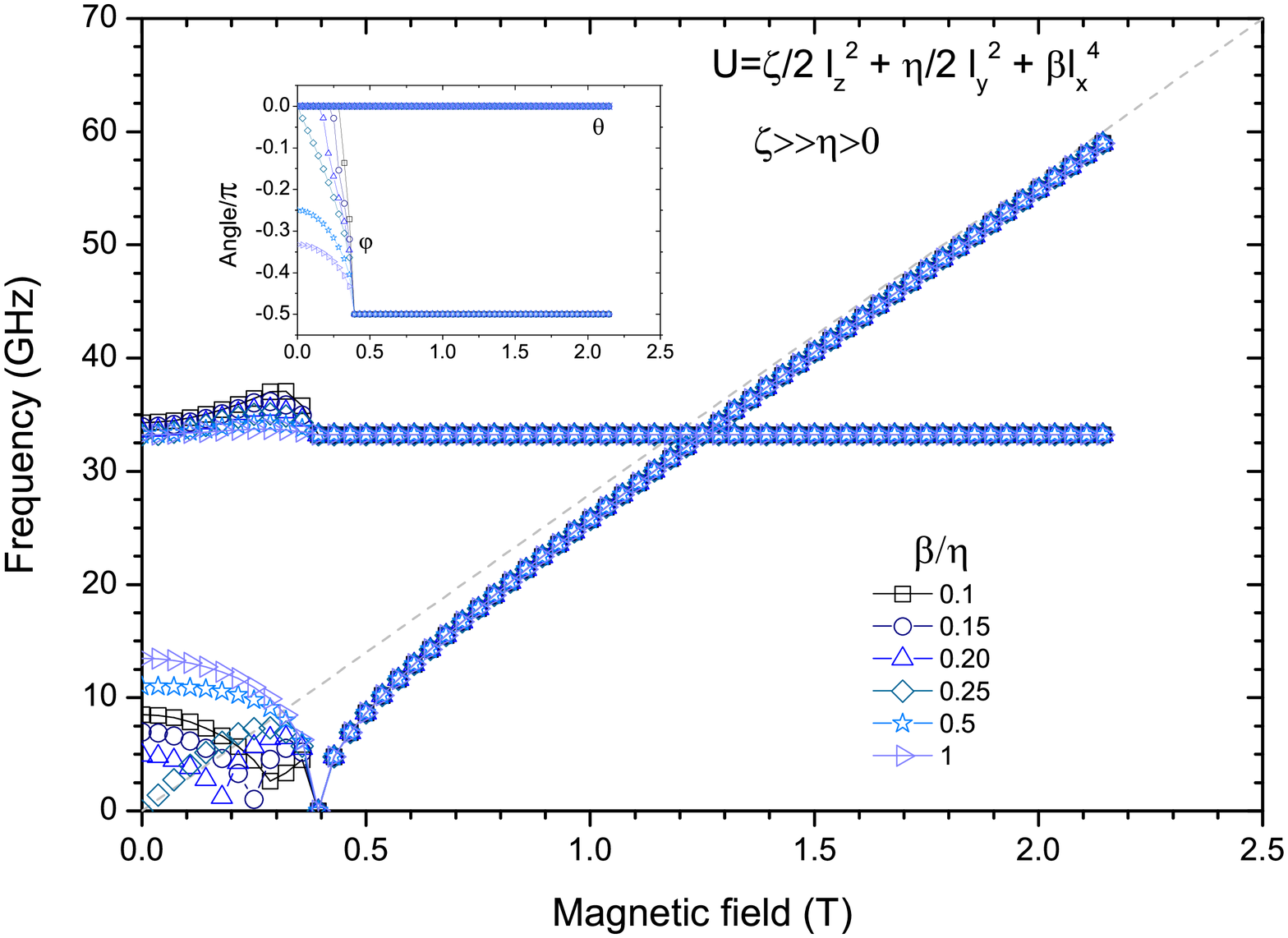}\\
  \caption{(Color online) Spectra of the model with a quartic term $\beta l_x^4$, ${\bf H}\parallel x$.
  Dashed line is paramagnetic resonance.}\label{FIG:minusbetax}
\end{figure}
\begin{figure}
    \includegraphics[width=0.5\textwidth]{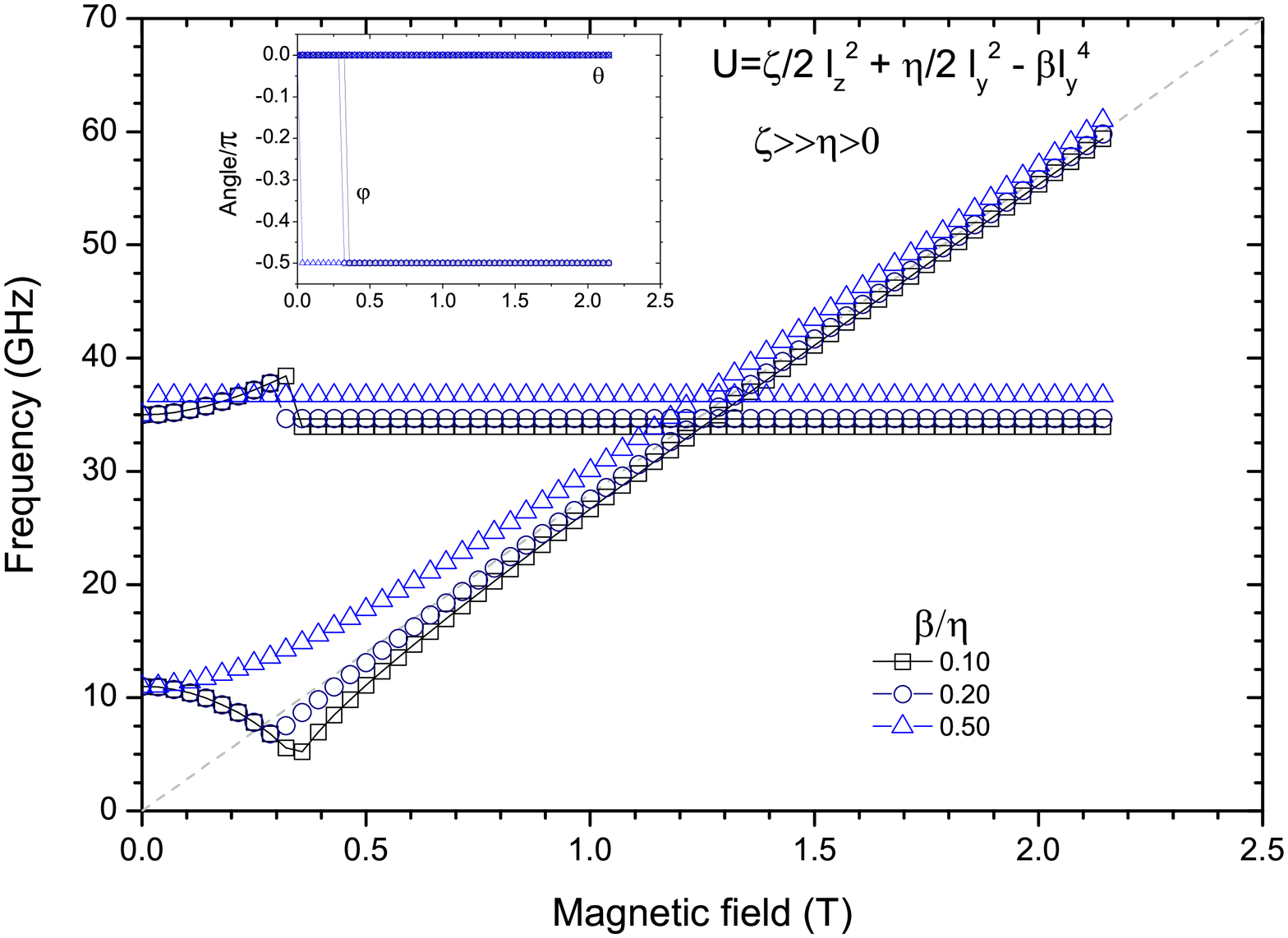}\\
  \caption{(Color online) Spectra of the model with a quartic term $-\beta l_y^4$, ${\bf H}\parallel x$.
  Dashed line is paramagnetic resonance.}\label{FIG:plusbetay}
\end{figure}
\begin{figure}
    \includegraphics[width=0.5\textwidth]{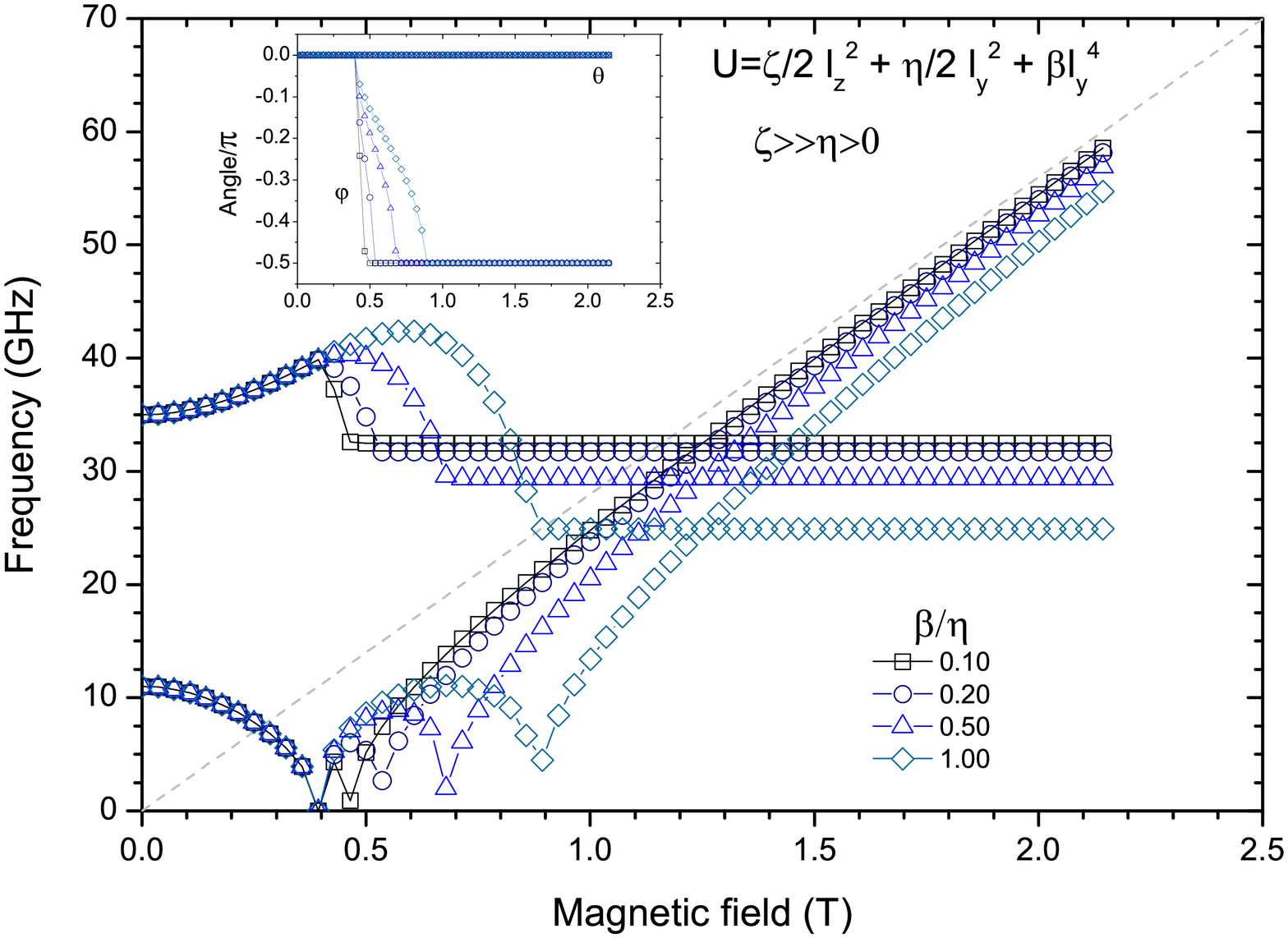}\\
  \caption{(Color online) Spectra of the model with a quartic term $\beta l_y^4$, ${\bf H}\parallel x$.
  Dashed line is paramagnetic resonance.}\label{FIG:minusbetay}
\end{figure}

Monoclinic symmetry allows for another second-order term, $l_{y}l_{z}$. Such a term results in a tilt of hard
and middle anisotropy axes, leaving easy axis undisturbed. In our experimental data, related to $xz$ plane
rotation of the magnetic field (Fig. \ref{FIG:RotTheta49GHz}), we do not notice any significant tilt of middle
axis from $c$ direction, and, therefore, we do not take $l_{y}l_{z}$ term into account.

We have to note, that anisotropic term $\eta l_{y}^{2}$ originates
from a weak orthorhombic distortion of a square lattice. Due to this
distortion, the lattice constants $b$ and $c$ differ in a relative
sense for about of ($\sim10^{-4}$). Hence, this term should be small
in comparison with, e.g. $\zeta l_{z}^2$, and can be comparable with
the contributions of higher order in components of $\mathbf{l}$.
There is a term $B(l_{x}l_{y})^{2}$ among the fourth-order terms,
allowed by symmetry for \Cupz. This term couples the components
$l_{x}, l_{y}$ and could result in the 'anisotropy reversal' as a
result of $\mathbf{l}$ reorientation. Indeed, considering a modified
Lagrange function
\begin{equation}
\begin{split}
\label{Lagr_quartic1} \mathcal{L}=
\frac{\chi_{\perp}}{\gamma^{2}}&\left(\frac{1}{2}\left(\dot{\mathbf{l}}+
{\gamma}[\mathbf{H}\times \mathbf{l}]\right)^{2}-{}\right.\\
-&
\left.\frac{(2\pi\Delta_{z})^{2}}{2}l_{z}^{2}-\frac{(2\pi\Delta_{y})^{2}}{2}l_{y}^{2}\right)+B(l_{x}l_{y})^{2},
\end{split}
\end{equation}
we obtain approximate frequencies, corresponding to in-plane
fluctuations of $\mathbf{l}$, for ground states before and after
reorientation:
$$\widetilde{\nu}(l\parallel
x)=\sqrt{\Delta_{y}^{2}+\dfrac{2B}{\chi_{\perp}}(\dfrac{g_{x}\mu_{B}}{2\pi\hbar})^{2}-(\dfrac{g_{x}\mu_{B}}{2\pi\hbar}
H)^{2}}$$ and
$$\widetilde{\nu}(l\parallel
 y)=\sqrt{-\Delta_{y}^{2}+\dfrac{2B}{\chi_{\perp}}(\dfrac{g_{x}\mu_{B}}{2\pi\hbar})^{2}+(\dfrac{g_{x}\mu_{B}}{2\pi\hbar}
H)^{2}}.$$ The field-independent constants under the square root signs in this relations may be treated by  use
of model (B2) and relations (A6,A7) as anisotropy constant for in-plane anisotropy (note that these constants
should be taken with the opposite signs). Thus, if $B>0$ is large enough, the in-plane anisotropy effectively
changes its sign. Significant value of $B$ might be provided by a magnetoelastic interaction. There are 13
elastic terms of the form of $u_{ik}u_{mn}$ and 20 magnetoelastic terms of the form of $u_{ik}l_{m}l_{n}$
allowed by symmetry for \Cupz\ [\onlinecite{LinesMagnetoelastic}]. From these terms we take for the
magnetoelastic contribution to potential energy (\ref{Epot_general}) the following terms:

\begin{equation}\label{EQtermME}
    E_{me}\propto u_{xy}l_{x}l_{y},
\end{equation}

and
\begin{equation}\label{EQtermEE}
    E_{ee}\propto\frac{ u_{xy}^{2}}{2}.
\end{equation}

These terms couple $l_{x}$ and $l_{y}$. Here $u_{ik}$ are components of the strain tensor.

 Minimization of energy, including described above magnetoelastic correction
(\ref{EQtermME},\ref{EQtermEE}), with respect to strain variable $u_{xy}$ will result in the magnetoelastic
correction in the form of $Bl_{x}^{2}l_{y}^{2}$.

We numerically calculate ground state and spectrum of the model (\ref{Lagr_quartic1}). For our numerical work we
choose parameters $\Delta_{z}=37$ and $\Delta_{y}=15$ GHz, and follow the perturbations, introduced by quartic
term $Bl_{x}^{2}l_{y}^{2}$. At Fig. \ref{FIG:Quartic1Joint} numerically calculated spectra for field along $x$
and different values of $B$ (namely,
$\dfrac{2B}{\chi_{\perp}}\left(\dfrac{g_{x}\mu_{B}}{2\pi\hbar\Delta_{y}}\right)^{2}=0,0.1,0.5,1,1.5,2$) are
present. Indeed, we can choose the value of $B$, which is large enough to push mode $\nu_{4}$ above paramagnetic
resonance, as it is clearly seen on Fig. \ref{FIG:Quartic1Joint}. But there remains a crucial difference between
the properties of model (\ref{Lagr_quartic1}) and experimental data, because the angular dependence is not
described. We found from ESR experiment that angular dependence of resonance line shows a step when switching
from $\nu_{a}$ to $\nu_{4}$, the anomalous mode $\nu_{a}$ exists in a narrow angle range near the $x$-axis, and
outside of this range all the data is perfectly described by a simple biaxial model. In contrast, the  model
with strong $B$ term (\ref{Lagr_quartic1}) shows smooth angular dependence (see insert on Fig.
\ref{FIG:Quartic1Joint}), which differs significantly from both simple biaxial model (\ref{Lagr_general}) and
experimental data (as on Fig. \ref{FIG:AnomalyAngularvarFQ}). Furthermore, as it is seen from Fig.
\ref{FIG:Quartic1Joint}, enhancement of $B$ term also increases a lower zero-field gap, while critical field of
spin-flop transition is still determined by $\Delta^{2}_{y}$ alone, as $l_{x}^{2}l_{y}^{2}$ combination gives
the same contribution to energies of both $\mathbf{l}\parallel x$ and $\mathbf{l}\parallel y$ phases. Therefore,
the magnetoelastic approach (\ref{Lagr_quartic1}) predicts the transposition of the antiferromagnetic resonance
frequency above the value of the paramagnetic resonance frequency, which is one of the manifestations of the
reversal of the in-plane anisotropy. Nevertheless, at the same time, the angular dependence of the resonance
field and relation between the gap and critical field do not correspond to the experiment even qualitatively.

Other related quartic terms also could not describe the observed
anomaly. We have analyzed in the same way the influence of
anisotropic terms $\pm \beta l_x^4$ and $\pm \beta l_y^4$. The
results of the calculation of the ground states (equilibrium values
of $\varphi,\theta$) and frequency-field dependencies are given in
Figs. \ref{FIG:plusbetax}, \ref{FIG:minusbetax},
\ref{FIG:plusbetay}, \ref{FIG:minusbetay}. One can see, that it is
impossible to find a value of $\beta$ which would correspond to the
observed pulling of the frequency above the paramagnetic resonance
frequency for $H>H_c$ along with the softening of the mode at
$H<H_c$, and with the valid relation $\gamma H_c=\Delta_y$.

\bibliography{Cupz}

\end{document}